\documentclass[fleqn,usenatbib]{mnras}

\usepackage{newtxtext}

\usepackage[T1]{fontenc}
\usepackage{ae,aecompl}
\usepackage[utf8]{inputenc}

\usepackage{amsmath}	
\usepackage{amssymb}	
\usepackage{mathtools}
\usepackage{graphicx}	
\usepackage[british]{babel}             
\usepackage{bm}
\usepackage{physics}
\usepackage{academicons}
\usepackage[dvipsnames]{xcolor}
\definecolor{orcidlogocol}{HTML}{A6CE39}
\newcommand{\orcid}[1]{\href{https://orcid.org/#1}{\textcolor[HTML]{A6CE39}{\aiOrcid}}}
\usepackage{booktabs}
\usepackage{ulem}
\usepackage{fontawesome}
\usepackage{extarrows}
\usepackage{mathrsfs}  
\usepackage{cuted}
\usepackage{tabularx,ragged2e}
\usepackage{makecell}

\newcommand{\rmcs}{{\text{cs}}}
\newcommand{\rmss}{{\text{ss}}}

\newcommand{\Nc}{{\langle N_{\text{c}} \rangle}}
\newcommand{\Ns}{{\langle N_{\text{s}} \rangle}}

\definecolor{lightseagreen}{rgb}{0.13, 0.7, 0.67}
\definecolor{palatinate}{rgb}{104, 36, 109}

\newcommand{\dndM}{{\frac{\dd{n}(M)}{\dd{M}}}}

\newcommand{\rmlos}{{\text{los}}}

\newcommand{\revision}[1]{\textcolor{black}{#1}}
\newcommand{\revisiontwo}[1]{\textcolor{black}{#1}}

\title[Halo Streaming Model]{A modern halo streaming model for redshift space distortions}

\author[C.-Z. Ruan et al.]{
Cheng-Zong Ruan$^{1,2}$\thanks{E-mail: cheng-zong.ruan@durham.ac.uk}, Baojiu Li$^{1}$\thanks{E-mail: baojiu.li@durham.ac.uk},  Carlton M. Baugh$^{1,3}$, Sownak Bose$^{1}$, \newauthor
Alexander Eggemeier$^{4}$ and David F. Mota$^{2}$ \\
$^{1}$Institute for Computational Cosmology, Department of Physics, Durham University, South Road, Durham DH1 3LE, UK\\
$^{2}$Institute of Theoretical Astrophysics, University of Oslo, 0315 Oslo, Norway\\
$^{3}$Institute for Data Science, Durham University, South Road, Durham DH1 3LE, UK\\
$^{4}$Universität Bonn, Argelander-Institut für Astronomie, Auf dem Hügel 71, 53121 Bonn, Germany
}

\date{Accepted XXX. Received YYY; in original form \today}

\pubyear{2026}

\begin{document}
\label{firstpage}
\pagerange{\pageref{firstpage}--\pageref{lastpage}}
\maketitle

\begin{abstract}
Accurate modelling of redshift-space distortions (RSD) in galaxy clustering is essential for extracting cosmological information from current and forthcoming large-scale structure surveys. While perturbation theory is reliable on large scales, much of the constraining power lies at intermediate and small separations, where nonlinear dynamics within and between dark matter haloes dominate.
We present a halo streaming model for nonlinear galaxy clustering in redshift space that is accurate and physically interpretable. 
Our framework combines the streaming model for RSD with a halo-model decomposition of the galaxy clustering into central/satellite and one-/two-halo contributions.
We build dedicated emulators for the key physical ingredients, trained on a suite of $N$-body simulations: halo mass functions, real-space halo two-point correlation functions, and pairwise velocity moments. 
By emulating these modular building blocks rather than the final redshift-space observable, this approach preserves physical transparency, enables targeted optimisation for each ingredient, and remains flexible to changes in tracer populations and galaxy-halo connection models.
The resulting halo streaming model reproduces the simulated nonlinear anisotropic clustering signal down to highly nonlinear scales, while achieving the computational efficiency required for cosmological parameter inference. 
This framework is designed to support full-shape RSD analyses for surveys such as DESI and \textit{Euclid}, facilitating precision measurements of structure growth and tests of gravity. 
All codes and trained emulators are publicly available in the \texttt{freyja} repository.\,\raisebox{-0.35\height}{\href{https://github.com/chzruan/freyja}{\includegraphics[scale=0.028]{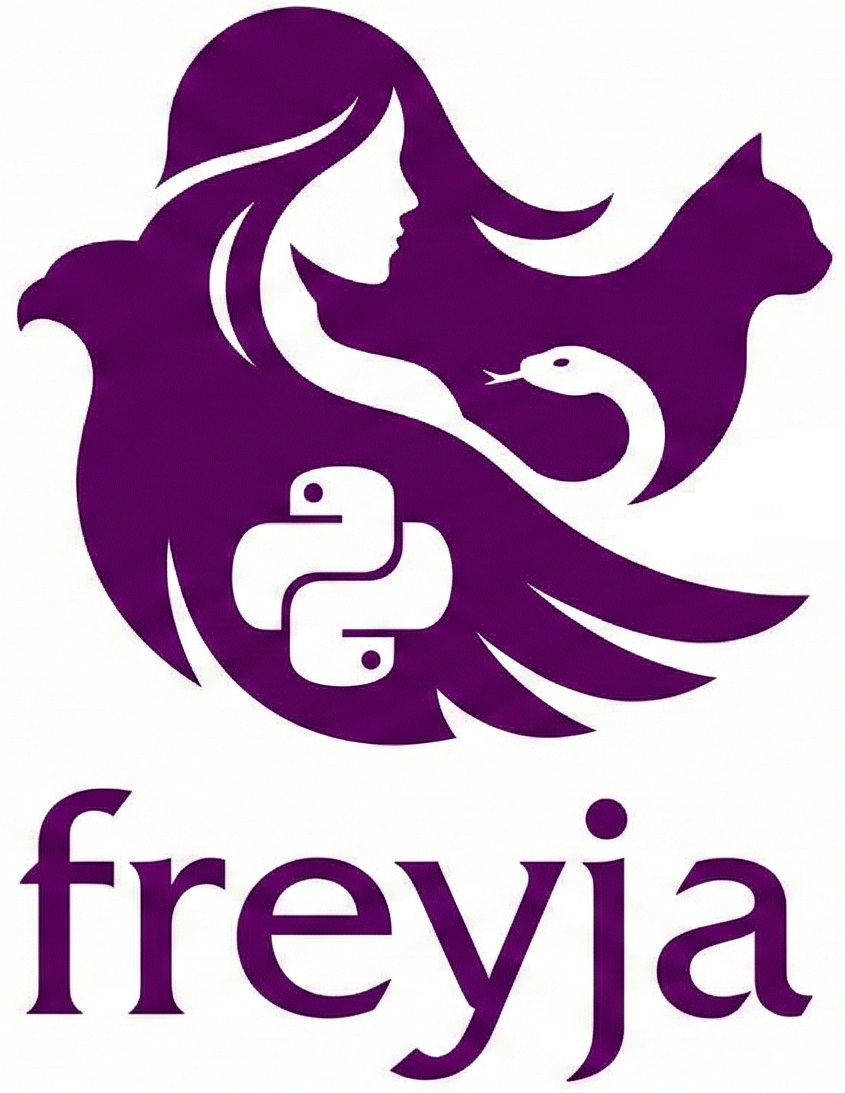}}}
\end{abstract}




\section{Introduction}
\label{sec:intro}

The present decade marks a golden era for extragalactic astronomy and cosmology. Large-scale structure (LSS) surveys, such as the Dark Energy Spectroscopic Instrument (DESI) \citep{DESI:2016arXiv161100036D} and the \textit{Euclid} mission \citep{Euclid:2011arXiv1110.3193L,AmendolaEuclid:2013LRR....16....6A,Euclid:2025A&A...697A...1E.Overview}, have begun to map the distribution of millions of galaxies over vast cosmic volumes. 
To fully exploit the statistical power of these surveys, theoretical predictions must match the sub-percent precision of the observational data.
A key probe for extracting cosmological information from spectroscopic surveys is redshift-space distortions of clustering \citep{Kaiser:1987MNRAS.227....1K}. The peculiar velocities of galaxies, mainly induced by the gravity of the inhomogeneous matter distribution, introduce an anisotropy in the observed clustering signal along the line of sight. 
This effect breaks the degeneracy between the galaxy bias and the growth rate of structure, $f$, making RSD a powerful tool for probing the growth history of the Universe \citep[e.g.][]{Peacock:2001Natur.410..169P.2dF,Hawkins:2003MNRAS.346...78H,Cole:2005MNRAS.362..505C.2dF,Guzzo:2008Natur.451..541G,2019MNRAS.485.2194H,Viglione:2026PhRvD.113l4017V}.

The streaming model \citep{Peebles:1980lssu.book.....P,Davis:1983ApJ...267..465D,Fisher:1995ApJ...448..494F,Scoccimarro:2004PhRvD..70h3007S.RSD,Kuruvilla:2018MNRAS.479.2256K} provides an accurate way of describing the RSD effect by decomposing the redshift-space clustering, $\xi^{\text{S}}(s_{\perp}, s_{\parallel})$, into two distinct ingredients: the real-space clustering of galaxies, $\xi^{\text{R}}(r)$, which encodes their spatial distribution, and the velocity information encoded by the pairwise line-of-sight velocity distribution $\mathcal{P}(v_{\parallel})$. 
As a result, the streaming model offers a clear physical interpretation of RSD and a convenient bridge between theoretical models, numerical simulations, and observational measurements.

The halo model \citep{Seljak:2000MNRAS.318..203S.HaloModel,White:2001MNRAS.321....1W,Cooray:2002PhR...372....1C,Schmidt:2016PhRvD..93f3512S.halomodel,Asgari2023OJAp....6E..39A.halomodelreview}, combined with the halo occupation distribution (HOD) framework \citep{Benson2000,Zheng:2004ApJ...610...61Z,Kravtsov:2004ApJ...609...35K}, provides a systematic and physically motivated description of galaxy clustering by connecting galaxies to their host dark matter haloes. 
It has been widely used to constrain cosmological and galaxy-halo connection parameters \citep[e.g.][]{Tinker:2005ApJ...631...41T,vandenBosch:2013MNRAS.430..725V,Mead:2021MNRAS.503.3095M,Mahony:2022MNRAS.515.2612M,Dvornik:2023A&A...675A.189D}.
In this approach, the full galaxy two-point correlation function is decomposed according to both galaxy type and halo environment. 
Galaxies are classified as centrals, residing at the centres of haloes, or satellites, orbiting within haloes.  Galaxy pairs are further separated into intra-halo and inter-halo contributions. 
This leads to a natural decomposition of the galaxy clustering signal into 1-halo and 2-halo terms, and each term receives contributions from central-central, central-satellite, and satellite-satellite pairs (see e.g. \citealt{Zheng:2005ApJ...633..791Z.HOD}), allowing different physical processes to be modelled in a modular and tailored way.

The unified \textit{halo streaming model} proposed  here integrates the halo- and streaming-model formalisms to decompose the redshift-space clustering signal into fundamental building blocks, following  \citet{Tinker:2007MNRAS.374..477T} (see also  \citet{Zheng:2016MNRAS.458.4015Z,Nishimichi:2019ApJ...884...29N,Hang:2022MNRAS.517..374H} and \citet{Cuesta-Lazaro:2023MNRAS.523.3219C}).
The halo model expresses the real-space galaxy clustering and velocity statistics as a sum of contributions from different galaxy types and halo configurations. 
The streaming model then maps each real-space component to redshift space.
This framework reduces the complexity of nonlinear redshift-space clustering to a collection of basic ingredients that can be modelled, calibrated, or emulated independently, providing a transparent and modular pathway from theoretical assumptions to observable galaxy clustering signals.

In recent years, the demand for precision has driven the field toward simulation-based approaches and emulators \citep[e.g.][]{Heitmann2010,DeRose:2019ApJ...875...69D.AemulusI, Zhai:2019ApJ...874...95Z,Miyatake:2021arXiv210100113M.DQ,Maksimova:2021MNRAS.508.4017M.ABACUSSUMMITOverview,Cuesta-Lazaro:10.1093/mnras/stae1234,Paillas:2023cpk}. After being trained using $N$-body simulations, emulators can predict halo and galaxy clustering statistics with a high accuracy. One strategy is to emulate the redshift-space observables directly, mapping the cosmological and galaxy-halo connection parameters straight onto the anisotropic correlation function or its multipoles. 
Such direct emulation is powerful by learning a high-dimensional mapping in which the individual physical ingredients are no longer separable: the resulting emulator might lack physical transparency, and a model trained on one tracer population or one set of cosmological assumptions may not generalise readily to different tracers or to extensions of the cosmological model.

The halo streaming model is a unified framework that combines the physical interpretability of analytical models with the accuracy of emulation. 
We adopt a \textit{modular emulation} strategy, modernizing the philosophy based on \citet{Tinker:2007MNRAS.374..477T}: rather than emulating the full redshift-space signal as a single object, we emulate only the physical building blocks, such as the halo mass function, the halo clustering statistics, and the halo pairwise velocity moments, then assemble them through analytic halo-streaming expressions.
This keeps the dimensionality of each emulator low, renders every ingredient interpretable in isolation, and allows physical intuition and external improvements to be brought to bear on each piece independently.
This work extends \citet{Cuesta-Lazaro:2023MNRAS.523.3219C}, which focuses on emulating the real-space clustering of haloes and combining this with a halo-model/HOD prescription, by providing the additional ingredients required by the streaming model, enabling direct predictions of the full anisotropic redshift-space clustering signal.

This paper is organised as follows. 
Section~\ref{sec:sim} describes the suite of $N$-body simulations used in this work.
In Section~\ref{sec:model}, we review the formalism of the halo streaming model and the decomposition of the galaxy correlation function.
Section~\ref{sec:emulators} presents the construction and tests of the emulators for the halo model ingredients.
Section~\ref{sec:mcmc} demonstrates that the model can accurately recover the cosmological parameters with mock galaxy observations.
Finally, we discuss the implications for future surveys and conclude in Section~\ref{sec:conclusion}. 
The code developed for this work, including the trained emulators, is made publicly available via the \texttt{freyja} repository\footnote{\url{https://github.com/chzruan/freyja}}.

\section{Simulations and mock catalogues}
\label{sec:sim}

\subsection{The \textsc{DEGRACE}-pilot simulation suite}
\label{subsec:HODs}

\begin{figure}
    \centering
    \includegraphics[width=\columnwidth]{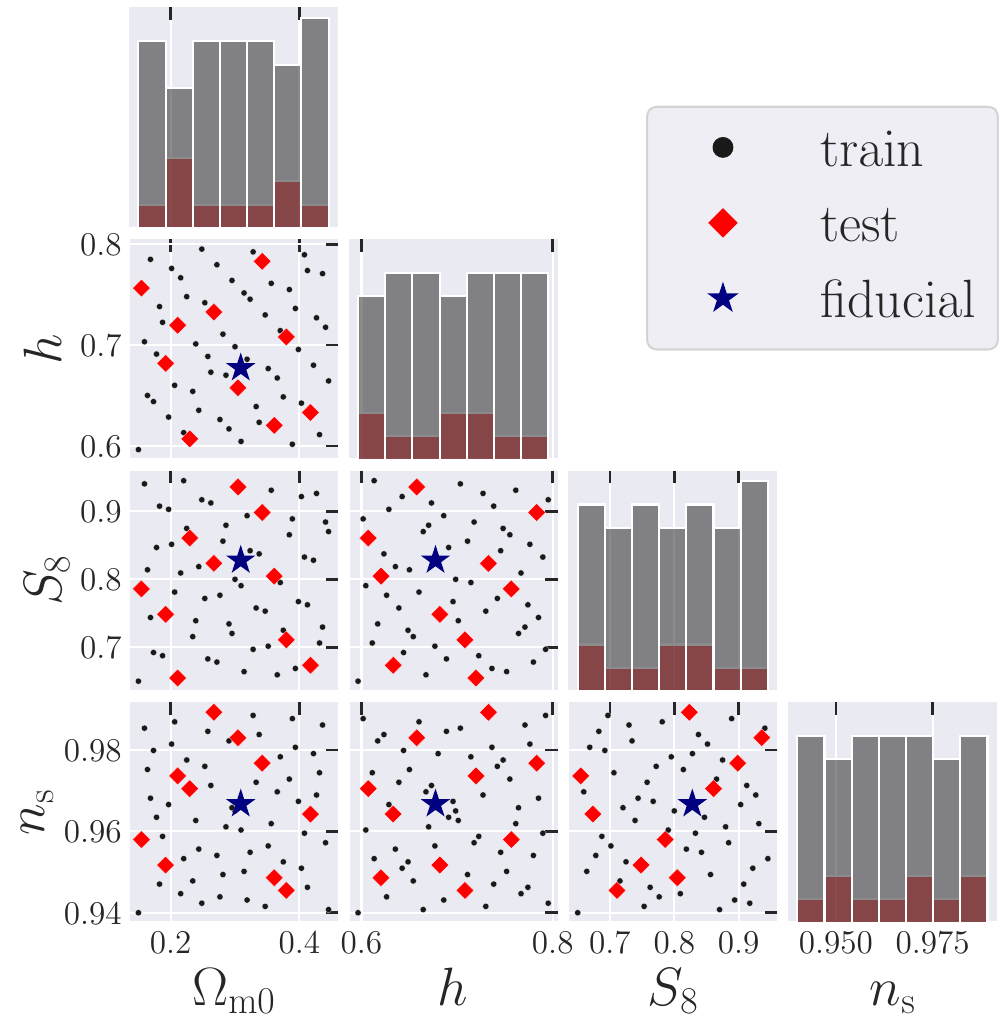}
    \caption{Visualisation of the cosmological parameter space, $\boldsymbol{\theta}_{\text{cos}}$, covered by the \textsc{DEGRACE}-pilot simulation suite. 
    The suite includes $64$ cosmological models, of which models $1$-$54$ are used for training and $55$-$64$ are reserved for testing. The star marks the fiducial \textit{Planck}-2015 $\Lambda$CDM cosmology fitted in the parameter-recovery test of Section~\ref{sec:mcmc} ($\Omega_{\text{m0}} = 0.3089$, $h = 0.6774$, $\sigma_8 = 0.8159$, $S_8 = 0.828$, $n_{\text{s}} = 0.9667$), which lies close to the centre of the design. \raisebox{-0.06in}{\href{https://github.com/chzruan/freyja/blob/main/paper_figs/cosmo_params.py}{\includegraphics[scale=0.023]{figs/freyja.jpg}}}
    }
    \label{fig:cosmoparams}
\end{figure}

To construct the ingredients of the halo streaming model and train the emulators, we make use of a new suite of $N$-body simulations, hereafter referred to as \textsc{DEGRACE}\footnote{Abbreviation for Dark Energy, GRavity and the Accelerated Cosmic Expansion.}-pilot. It comprises $64$ $\Lambda$CDM models that span a four-dimensional cosmological parameter space,
\begin{align}
    \boldsymbol{\theta}_{\text{cos}} = \Big\{\Omega_{\text{m0}},\, h,\, S_8,\, n_{\text{s}}\Big\},
    \label{eqn:cos_param_space}
\end{align}
where $h \equiv H_0 / (100\,\mathrm{km}\,\text{s}^{-1}\,\mathrm{Mpc}^{-1})$ with $H_0$ being the Hubble constant; $S_8 \equiv \sigma_8 \sqrt{\Omega_{\mathrm{m0}} / 0.3}$ characterises the amplitude of matter fluctuations; $\Omega_{\textrm{m}0}$ is the (baryonic plus cold dark) matter density parameter today; and $n_{\text{s}}$ denotes the spectral index of the primordial power spectrum.  
For each cosmological model, we generate five independent realisations in order to suppress sample variance. 
We employ a Sobol sequence \citep{SOBOL:1571135650577277824,Matousek:1998527SobolSequence} to sample the parameter values, providing an efficient and statistically uniform coverage of the multidimensional space, as illustrated in Fig.~\ref{fig:cosmoparams}. 
The Sobol design is similar to Latin hypercube sampling, but further allows for straightforward extensions of the sampling.
The parameter ranges explored are
\begin{equation}
\begin{split}
    0.15 &< \Omega_{{\text{m0}}} < 0.45,  \\
    0.60 &< h < 0.80,               \\
    0.65 &< S_8 < 0.95,             \\
    0.94 &< n_{\text{s}} < 0.99.       \\
\end{split}
\label{eqn:param_range_slhc}
\end{equation}
The physical baryonic and neutrino densities are fixed to $\Omega_{\text{b0}}h^2 = 0.0224$ and $\Omega_{\nu0} h^2 = 0.00064$, respectively, corresponding to a total neutrino mass of $0.06\,\mathrm{eV}$. Assuming spatial flatness, the dark energy density parameter is given by $\Omega_{\Lambda0} = 1 - \Omega_{{\text{m0}}}$.  
In addition to $\Lambda$CDM, we have also generated suites of modified-gravity simulations based on $f(R)$ and DGP models; however, the present work focuses exclusively on the standard cosmological model, leaving the analysis of the MG runs to future studies.

All simulations were performed with the \href{https://www.skiesanduniverses.org/Simulations/GLAM/}{\textsc{GLAM}} code \citep{Klypin:2017iwu,Klypin:2020tud}, a particle-mesh $N$-body solver optimised for large ensembles of simulations. 
\textsc{GLAM} employs \textsc{OpenMP}-based shared-memory parallelism, allowing many independent realisations to be run across multiple compute nodes. 
Combined with its fixed force and mass resolutions, this design makes \textsc{GLAM} particularly well suited for large emulator campaigns that require a high throughput of simulations spanning a wide range of cosmological models. 
Its speed and scalability enable the efficient production of halo catalogues and mock galaxy samples, which form the essential training data for emulator construction.

Each run evolves $N_{\text{p}} = 2048^3$ particles within a cubic volume of side length $L = 1024\,h^{-1}\,\mathrm{Mpc}$. The gravitational potential is computed on a mesh of $N_{\text{g}} = 4096^3$ cells, corresponding to a spatial resolution of $0.25\,h^{-1}\,\mathrm{Mpc}$. 
This translates to a minimum reliably resolved halo mass of $10^{12.4}\,h^{-1}M_{\odot}$ at redshift $z = 0.25$ for $\Omega_{\text{m0}} \approx 0.3$ \citep[cf.][]{Hern:2021MNRAS.503.2318H,Ruan:2022JCAP...05..018R}.
The simulation configuration provides a good compromise between large volume (required for capturing long-wavelength modes and suppressing sample variance) and sufficient mass resolution to resolve dark matter haloes hosting galaxies targeted by ongoing and forthcoming surveys.

In \textsc{DEGRACE}-pilot, the minimum reliably resolved halo mass of $M = 10^{12.4}\,h^{-1}M_{\odot}$ (corresponding to $\approx 250$
particles) is well suited to the massive-halo tracers targeted by current spectroscopic surveys. Recent DESI One-Percent HOD fits for luminous red galaxies (LRGs) give central turn-on masses of $\log_{10}[M_{\text{cut}}/(h^{-1}M_{\odot})]=12.79^{+0.15}_{-0.07}$ and $12.64^{+0.17}_{-0.05}$ in the fiducial $0.4 < z < 0.6$ and $0.6 < z < 0.8$ bins, respectively, with corresponding mean host-halo masses $\log_{10}[\bar M_{\text{h}}/(h^{-1}M_{\odot})] = 13.40$ and $13.24$ \citep{Yuan:2024MNRAS.530..947Y}. 
These values imply that a minimal, non-conservative mass-resolution requirement for DESI-like LRG HOD modelling is $M \sim 10^{12.6}\,h^{-1}M_{\odot}$, set roughly by the lowest DESI LRG $M_{\text{cut}}$, whereas a conservative choice that resolves the low-occupation tail of the central HOD is $M \lesssim 10^{12.0}$--$10^{12.3}\,h^{-1}M_{\odot}$, especially when including the broader high-redshift LRG occupation at $z\simeq1$. 
An even more conservative \textsc{Abacus}-like reference scale would be $M \lesssim 2 \times 10^{11}\,h^{-1}M_{\odot}$, comparable to the $100$-particle reliability threshold quoted for the \textsc{AbacusSummit} light-cone halo catalogues \citep{Hadzhiyska:2022MNRAS.509.2194H}. 
The \textsc{DEGRACE}-pilot threshold is therefore below the characteristic DESI LRG central-occupation scale and far below the mean LRG host mass, so the present suite is adequate for LRG-like analyses, although it should not be regarded as a fully conservative choice for analyses explicitly sensitive to the lowest-mass LRG occupation tail. 
This conclusion is consistent with previous LRG samples: eBOSS LRGs have an inferred mean host mass $\simeq 2.5 \times 10^{13}\,h^{-1}M_{\odot}$, while BOSS LOWZ and CMASS have mean host masses $\simeq 5.2 \times 10^{13}$ and $\simeq 2.6 \times 10^{13}\,h^{-1}M_{\odot}$, respectively \citep{Reid:2016MNRAS.455.1553R,Zhai:2017ApJ...848...76Z}. 
It is, however, insufficient at the low-mass end for fainter tracers such as the DESI Bright Galaxy Survey (BGS), emission-line galaxies (ELGs), or the \textit{Euclid} spectroscopic sample, whose host haloes extend below the converged range: DESI BGS mock construction requires HODs down to faint absolute-magnitude thresholds, DESI ELG HOD fits give $\log_{10}\langle M \rangle \simeq 11.9$, and \textit{Euclid} spectroscopic mocks are designed to resolve substantially lower-mass haloes hosting emission-line galaxies \citep{Rocher:2023JCAP...10..016R,Smith:2024MNRAS.532..903S.BGSMocks,Euclid:2025A&A...704A.306E}.
We therefore do not advocate aggressive extrapolation of the emulated halo statistics below $10^{12.4}\,h^{-1}M_{\odot}$. 
A higher-resolution simulation suite that extends the reliable halo-mass range to lower masses is currently being constructed within the same \texttt{freyja} pipeline, and will support ELG- and BGS-targeted analyses in follow-up work.

\subsection{HOD prescriptions}
\label{subsec:HODs}

The halo streaming model can be combined with arbitrary halo occupation distribution prescriptions to model galaxy clustering. In this work, we adopt a standard HOD parameterisation as a baseline to test and validate our model pipeline.

Galaxy populations are assigned to dark matter haloes using the conventional HOD framework \citep{Berlind:2003ApJ...593....1B,Zheng:2005ApJ...633..791Z.HOD}. The free parameters of the model are
\begin{align}
    \boldsymbol{\theta}_{\text{HOD}} = \Big\{\log M_{\text{cut}},\, \log M_1,\, \alpha,\, \sigma,\, \alpha_{\text{c}}^{\text{vel}},\, \alpha_{\text{s}}^{\text{vel}} \Big\},
\end{align}
where $\log M_{\text{cut}}$ and $\sigma$ control the abundance of central galaxies, $\log M_1$ and $\alpha$ describe the satellite occupation, and $\alpha_{\text{c}}^{\text{vel}}$ and $\alpha_{\text{s}}^{\text{vel}}$ are velocity bias parameters for centrals and satellites, respectively. The parameter $\kappa$, which shifts the minimum mass scale for satellite occupation, is fixed to unity since it has a negligible impact on the clustering statistics considered here.

The mean numbers of central and satellite galaxies hosted by a halo of mass $M$ are specified as
\begin{align}
   \Nc(M) &= \frac{1}{2} \qty[ 1 + \mathrm{erf}\qty( \frac{\log M - \log M_{\text{cut}}}{\sqrt{2}\,\sigma} ) ], \\
    \Ns(M) &=\Nc(M)\,\qty(\frac{M - \kappa M_{\text{cut}}}{M_1})^\alpha,
\end{align}
where $\Nc(M)$ transitions smoothly from $0$ to $1$, while $\Ns$ scales as a power law above the mass threshold $\kappa M_{\text{cut}}$.  
Central galaxies are placed at the centres of their host haloes. Satellite galaxies are distributed radially according to the host halo density profile, with isotropic angular positions. 
\begin{table}
\centering
\caption{HOD parameter values used to generate the LOWZ-like (LRG) mock galaxy catalogue analysed in this work, following the Zheng07-style parameterisation of Section~\ref{subsec:HODs}. Masses are in units of $h^{-1}M_{\odot}$. The shift parameter $\kappa$ is fixed to unity (Section~\ref{subsec:HODs}). $\sigma$ is the (linear) scatter in the central-occupation error function, and $\alpha_{\text{c}}^{\text{vel}}$, $\alpha_{\text{s}}^{\text{vel}}$ are the central and satellite velocity-bias parameters.}
\label{tab:hod_params}
\renewcommand{\arraystretch}{1.3}
\begin{tabular}{@{} l c c @{}}
\toprule
\revision{Parameter} & \revision{Symbol} & \revision{Value} \\
\midrule
\revision{Central cutoff mass}        & \revision{$\log_{10} M_{\text{cut}}$}            & \revision{$13.62$}  \\
\revision{Satellite normalisation mass} & \revision{$\log_{10} M_1$}                       & \revision{$14.42$}  \\
\revision{Central scatter}             & \revision{$\sigma$}                               & \revision{$0.6915$} \\
\revision{Satellite slope}             & \revision{$\alpha$}                               & \revision{$0.9168$} \\
\revision{Shift parameter}             & \revision{$\kappa$}                               & \revision{$1.0$}    \\
\revision{Central velocity bias}       & \revision{$\alpha_{\text{c}}^{\text{vel}}$}      & \revision{$0.8$}    \\
\revision{Satellite velocity bias}     & \revision{$\alpha_{\text{s}}^{\text{vel}}$}      & \revision{$1.2$}    \\
\bottomrule
\end{tabular}
\end{table}\revision{The HOD parameter values used to generate the mock galaxy catalogue analysed in this work (a LOWZ-like LRG sample) are listed in Table~\ref{tab:hod_params}.}

The velocity of each galaxy is given by the halo bulk velocity plus a stochastic component along the Cartesian coordinates:
\begin{align}
    \boldsymbol{v}_{\text{c}} &= \boldsymbol{v}_{\text{h}} + \boldsymbol{v}_{\text{c},\text{sto}}, \\
    \boldsymbol{v}_{\text{s}} &= \boldsymbol{v}_{\text{h}} + \boldsymbol{v}_{\text{s},\text{sto}},
\end{align}
where the stochastic terms are drawn from independent Gaussian distributions,
\begin{align}
    v^{x/y/z}_{\text{c}/\text{s}, \text{sto}} \sim \mathcal{N}\!\left(0,\, \alpha^2_{\text{c}/\text{s}}\, \sigma_{\text{vir}}^2(M_{\text{h}}) \right).
\end{align}
Here $\sigma_{\text{vir}}(M_{\text{h}})$ denotes the halo virial velocity dispersion. \revision{In this work we measure $\sigma_{\text{vir}}(M_{\text{h}})$ and fit directly from the member-particle distribution within $R_{\text{h}}$ for each halo. The analytic virial estimate}
\begin{align}
    \sigma_{\text{vir}}(M_{\text{h}}) \equiv \sqrt{\frac{G M_{\text{h}}}{2R_{\text{h}}}}
\end{align}
\revision{or its extensions can be used in semi-analytic predictions in which particle-level information is unavailable.}
The velocity bias parameters $\alpha^{\text{vel}}_{\text{c}}$ and $\alpha^{\text{vel}}_{\text{s}}$ allow for deviations from this virial expectation and are therefore essential for accurately modelling redshift-space clustering. 
While a variety of galaxy-halo connection prescriptions have been explored in the literature \citep[e.g.][]{Zheng:2005ApJ...633..791Z.HOD,Zheng:2007ApJ...667..760Z}, our implementation places particular emphasis on the velocity assignment scheme. This choice directly determines the line-of-sight pairwise velocity distributions of galaxies, which in turn control the fidelity of RSD modelling within the streaming framework.

\subsection{HOD Variants and Extensions}
\label{subsec:HOD_variants}

A wide range of extensions to the standard halo occupation distribution framework have been proposed to capture increasingly complex aspects of the galaxy-halo connection. These include models incorporating galaxy assembly bias, where galaxy occupation depends on secondary halo properties other than mass, as well as alternative formulations such as the conditional luminosity or stellar mass functions. More recent approaches combine empirical and physical modelling, for example through subhalo abundance matching \citep[SHAM,][]{Conroy:2006ApJ...647..201C.SHAM} or hybrid HOD-SHAM schemes.
\revision{Concrete examples include the conditional abundance-matching framework of \citet{Reddick:2013ApJ...771...30R}, in which central galaxies follow an HOD-like prescription while satellites are assigned by subhalo abundance matching, and the \textsc{AbacusHOD} implementation of \citet{Yuan:2022MNRAS.510.3301Y.ABACUSHOD}, which augments the standard \citet{Zheng:2005ApJ...633..791Z.HOD} HOD with subhalo and assembly-bias modules.} While such variants can reproduce specific observational features with high fidelity, they typically introduce additional free parameters and degeneracies that complicate parameter inference and model calibration.

Here, we deliberately adopt a minimal, mass-only HOD based on the standard parameterisation \citep{Zheng:2005ApJ...633..791Z.HOD}, without assembly bias \citep{Hearin:2016MNRAS.460.2552H.assemblybias}. 
Satellite occupations are assumed to follow Poisson statistics, galaxy positions trace spherical NFW profiles, and galaxy velocities are modelled with Gaussian stochastic components augmented by central and satellite velocity bias parameters. 
This baseline choice is sufficient for testing the halo streaming framework and emulator pipeline developed here, while maintaining a clear separation between cosmological and astrophysical degrees of freedom. A more comprehensive exploration of extended HOD prescriptions is beyond the scope of this work.

\section{The halo streaming model of redshift-space clustering}
\label{sec:model}

\subsection{The streaming model for RSD}
\label{subsec:streaming_model_rsd_sekrjfsd}

Within the halo model framework, the galaxy-galaxy two-point correlation function, $\xi_{\text{gg}}$, can be decomposed into its constituent one-halo and two-halo terms. This decomposition is valid in both real and redshift space and is given by:
\begin{align}
    \xi_{\text{gg}} = f_{\text{c}}^2 \xi_{\mathrm{cc}}^{\text{2h}} + 2f_{\text{c}} f_{\text{s}} \qty(\xi_{\mathrm{cs}}^{\text{1h}} + \xi_{\mathrm{cs}}^{\text{2h}}) + f_{\text{s}}^2 \qty(\xi_{\mathrm{ss}}^{\text{1h}} + \xi_{\mathrm{ss}}^{\text{2h}}), \label{eqn:xigg_hm_hod_decom_ewrfsd}
\end{align}
where $f_{\text{c}} \equiv n_{\text{c}} / n_{\text{g}}$ and $f_{\text{s}} \equiv n_{\text{s}} / n_{\text{g}}$ are the number fractions of central and satellite galaxies, respectively, and $n_{\textrm{c}}$, $n_{\textrm{s}}$ and $n_{\textrm{g}}$ are the numbers of central, satellite and all galaxies.

To model RSD, we need to consider the kinematics of galaxy pairs. For a pair of tracer galaxies with real-space positions $\boldsymbol{r}_1, \boldsymbol{r}_2$ and peculiar velocities $\boldsymbol{v}_1, \boldsymbol{v}_2$, their separation vector is $\boldsymbol{r} \equiv \boldsymbol{r}_2 - \boldsymbol{r}_1$ and their pairwise peculiar velocity is $\boldsymbol{v} \equiv \boldsymbol{v}_2 - \boldsymbol{v}_1$.
The redshift-space clustering is distorted by the component of the pairwise velocity along the line-of-sight (LOS). To describe this effect, we adopt a coordinate system in which the LOS direction $\hat{\boldsymbol{\ell}}$ is fixed along the $z$-axis under the plane-parallel approximation, i.e., $\hat{\boldsymbol{\ell}} \approx \hat{\boldsymbol{z}}$. It is useful to decompose the pairwise velocity $\boldsymbol{v}$ into components that are radial ($v_{\mathrm{r}}$) and transverse ($v_{\mathrm{t}}$) with respect to the pair separation vector $\boldsymbol{r}$. These components are defined by projecting $\boldsymbol{v}$ onto the unit vectors:
\begin{align}
    \hat{\boldsymbol{r}} \equiv \boldsymbol{r} / r, \quad \hat{\boldsymbol{t}} \equiv \hat{\boldsymbol{r}} \times (\hat{\boldsymbol{r}} \times \hat{\boldsymbol{\ell}}).
\end{align}
The radial, transverse, and LOS components of the pairwise velocity are then related by:
\begin{align}
    v_{\text{r}} &= \boldsymbol{v} \cdot \hat{\boldsymbol{r}}, \quad v_{\text{t}} = \boldsymbol{v} \cdot \hat{\boldsymbol{t}}, \\
    v_{\parallel} &= \boldsymbol{v} \cdot \hat{\boldsymbol{\ell}} = v_{\text{r}} \cos\theta + v_{\text{t}} \sin\theta, \label{eqn:vlos_and_vrt_rewfds}
\end{align}
where $\theta$ is the angle between the separation vector $\boldsymbol{r} = (r_{\perp}, r_{\parallel})$ and the LOS direction.

The LOS peculiar velocity distorts the observed separation of the galaxy pair. 
The mapping from the real-space separation vector $\boldsymbol{r}$ to the redshift-space separation vector $\boldsymbol{s}$ is given by:
\begin{align}
    \boldsymbol{s} = \boldsymbol{r} + \frac{v_{\parallel}}{aH(a)} \hat{\boldsymbol{z}},
\end{align}
where $a$ is the scale factor and $H(a)$ is the Hubble parameter. This mapping forms the basis of the streaming model \citep[e.g.][]{Peebles:1980lssu.book.....P,Fisher:1995ApJ...448..494F,Scoccimarro:2004PhRvD..70h3007S.RSD,Kuruvilla:2018MNRAS.479.2256K,Kuruvilla:2020JCAP...07..043K}, which relates the redshift-space 2PCF, $\xi^{\text{S}}(s_{\perp}, s_{\parallel})$, to its real-space counterpart, $\xi^{\text{R}}(r)$:
\begin{align}
    1 + \xi^{\text{S}}(s_{\perp}, s_{\parallel}) = \int_{-\infty}^{\infty} \mathrm{d}{r_{\parallel}} \qty[1 + \xi^{\text{R}}(r)]\, \mathcal{P}\qty(v_{\parallel} | r_{\perp}, r_{\parallel}).
\end{align}
In this equation, the respective separation components are related by $v_{\parallel} = (s_{\parallel} - r_{\parallel})/(aH)$, $r_{\perp} = s_{\perp}$, and $r = \sqrt{r_{\perp}^2 + r_{\parallel}^2}$.
The kernel of the integral, $\mathcal{P}\big(v_{\parallel} | \boldsymbol{r} \big)$, is the probability distribution function (PDF) of the pairwise LOS velocity for a given real-space separation $\boldsymbol{r} = (r_{\perp}, r_{\parallel})$.

The main idea here is to use the halo model together with the HOD framework to explicitly separate galaxy pairs into one-halo (pairs within the same halo) and two-halo (pairs in different haloes) components, as well as into central-central, central-satellite, and satellite-satellite pair types:
\begin{align}
\xi_{\text{gg}}^{\text{S}}(s_{\perp}, s_{\parallel})
\xrightarrow[\text{HOD}]{\text{halo model}}\left\{\begin{lgathered}
    \revision{\xi_{\text{cs}}^{\text{S,1h}}(r), \xi_{\text{ss}}^{\text{S,1h}}(r)} \\
    \revision{\xi_{\text{cc}}^{\text{S,2h}}(r), \xi_{\text{cs}}^{\text{S,2h}}(r), \xi_{\text{ss}}^{\text{S,2h}}(r)}.
\end{lgathered}
\right.
\end{align}
We then apply the streaming model to each pair type separately, so that the redshift-space 2PCF can be expressed in terms of its real-space clustering and pairwise velocity distribution components,
\begin{align}
\revision{\xi^{\text{S,2h}}_{\text{cc}}}(s_{\perp}, s_{\parallel})
\xrightarrow{\text{streaming model}}
\left\{
\begin{lgathered}
    \revision{\xi^{\text{R,2h}}_{\text{cc}}}(r), \\
    \mathcal{P}^{\text{2h}}_{\text{cc}}(v_{\parallel} | \bm{r}).
\end{lgathered}
\right.
\end{align}

\subsection{One-halo terms}
\label{subsec:xi_R_1h_halomod}

In this section, we present the analytical expressions for the one-halo terms of the 2PCF, which describe the correlation of galaxy pairs residing within the same dark matter halo. 
The real-space one-halo terms for central-satellite (cs) and satellite-satellite (ss) pairs are given by:
\begin{align}
    \xi^{\text{R,1h}}_{\text{cs}}(r) &\equiv \frac{1}{n_{\text{c}} n_{\text{s}}}\int\dd{M}\dndM \langle N_{\text{c}} N_{\text{s}}\rangle(M) \, u_{\text{s}}(r|M), \label{eqn:xi_R_1h_cs_dsfgsr} \\
    \xi^{\text{R,1h}}_{\text{ss}}(r) &= \frac{1}{n_{\text{s}}^2} \int\dd{M}\dndM \langle N_{\text{s}} (N_{\text{s}} - 1)\rangle (M) \lambda_{\text{ss}}(r|M). \label{eqn:xi_R_1h_ss_dsfgsr}
\end{align}
Here, $\dd{n}/\dd{M}$ is the halo mass function. 
The term $u_{\text{s}}(r|M)$ is the normalised radial number density profile of satellite galaxies within a halo of mass $M$. The term $\lambda_{\text{ss}}(r|M)$ is the auto-convolution of the satellite profile:
\begin{align}
    \lambda_{\text{ss}}(r|M) \equiv \int\dd{\boldsymbol{x}} u_{\text{s}} (\boldsymbol{x} | M) \, u_{\text{s}} (\boldsymbol{x} + \boldsymbol{r} | M).
\end{align}
The terms $\langle N_{\text{c}} N_{\text{s}}\rangle (M)$ and $\langle N_{\text{s}} (N_{\text{s}} - 1)\rangle (M)$ are the first and second factorial moments of the HOD, representing the average number of central-satellite pairs and satellite-satellite pairs, respectively. We adopt the standard HOD prescription for luminous red galaxies: a halo hosts at most one central galaxy, and the number of satellites, $N_{\text{s}}$, follows a Poisson distribution. 
Under these assumptions, these terms simplify to \citep[cf.~Equations~(4)-(6) of][]{Zheng:2005ApJ...633..791Z.HOD}:
\begin{align}
    \expval{N_{\text{c}} N_{\text{s}}}\!(M) &= \expval{N_{\text{s}}}\!(M) \\
    \expval{N_{\text{s}} (N_{\text{s}} - 1)}\!(M) &= \big[\!\expval{N_{\text{s}}}\!(M)\big]^2.
\end{align}

To model the one-halo term in redshift space, we must account for the internal (virial) motions of galaxies within haloes. For a galaxy pair in the same halo, the bulk velocity of the halo cancels out, and their relative peculiar velocity is determined by their orbits. Following the velocity assignment scheme in Section~\ref{subsec:HODs}, we model the pairwise LOS velocity distributions, $\mathcal{P}(v_{\parallel})$, as Gaussians with zero mean. The variances depend on the halo velocity dispersion, $\sigma(M)$, and the velocity bias parameters, $\alpha^{\text{vel}}$:
\begin{align}
    \mathcal{P}_{\rmcs}^{\text{1h}} (v_{\parallel} | M)
    &\sim
    \mathcal{N}\qty[
        0,
        \qty(\alpha_{\text{c}}^{\text{vel}} \sigma(M))^2
        +
        \qty(\alpha_{\text{s}}^{\text{vel}} \sigma(M))^2
    ], \\
    \mathcal{P}_{\rmss}^{\text{1h}} (v_{\parallel} | M)
    &\sim
    \mathcal{N}\qty[
        0,
        \qty(\sqrt{2} \alpha_{\text{s}}^{\text{vel}} \sigma(M))^2
    ],
\end{align}
where $\mathcal{N}[\mu, \sigma^2]$ denotes a Gaussian distribution with mean $\mu$ and variance $\sigma^2$.

\revision{This Gaussian description can be viewed as the moment-level analogue of the particle-based satellite velocity-bias prescription \citep[e.g.][]{Yuan:2024MNRAS.530..947Y}, which can be expressed as
\begin{equation}
    {\boldsymbol{v}}_{\text{sat}}
    =
    {\boldsymbol{v}}_{\text{h}}
    +
    \alpha_{\text{sat}}^{\text{vel}}
    \left(
    {\boldsymbol{v}}_{\text{particle}}
    -
    {\boldsymbol{v}}_{\text{h}}
    \right).
\end{equation}
This prescription amounts to sampling the empirical internal velocity distribution of each halo. To leading order, however, the one-halo redshift-space clustering is primarily sensitive to the second moment of this distribution. For relaxed haloes, whose internal velocities are approximately Maxwellian, replacing the particle draw by a Gaussian with dispersion $\alpha_{\text{sat}}^{\text{vel}}\sigma(M)$ therefore captures the same phenomenology, while the velocity-bias parameter absorbs residual differences between the adopted velocity model and the true internal velocity distribution.}

\revision{The main modelling choice is therefore the halo velocity dispersion $\sigma(M)$. The modular design of the \texttt{freyja} framework allows this quantity to be specified in several ways. For example, one may use the velocity dispersion stored in the halo catalogue by the halo finder, adopt calibrated virial scaling relations \citep{Evrard:2008ApJ...672..122E}, or, assuming that satellite radii are drawn from an NFW profile, solve the isotropic Jeans equation to obtain the local velocity dispersion $\sigma(r|M)$ at the assigned satellite radius.}

Applying the streaming model of Section~\ref{subsec:streaming_model_rsd_sekrjfsd}, we convolve the real-space profiles with these velocity PDFs to obtain the redshift-space one-halo terms:
\begin{align}
    1 + \xi^{\text{S,1h}}_{\text{cs}} (s_{\perp}, s_{\parallel}) &= \frac{1}{n_{\text{c}} n_{\text{s}}}   \int\dd{M} \dndM  \Ns(M) \cdots\notag \\
    &\phantom{= \frac{1}{n_{\text{c}} n_{\text{s}}} } \int  \dd{r_{\parallel}}  u_{\text{s}}(r | M) \, \mathcal{P}^{\text{1h}}_{\text{cs}} \qty(s_{\parallel} - r_{\parallel}  \big| M), \\
    1 + \xi^{\text{S,1h}}_{\text{ss}} (s_{\perp}, s_{\parallel}) &= \frac{1}{n_{\text{s}}^2} \int\dd{M} \dndM  \qty[\Ns(M)]^2  \cdots \notag \\
    &\phantom{= \frac{1}{n_{\text{s}}^2}}  \int \dd{r_{\parallel}} \lambda_{\text{ss}}(r|M) \mathcal{P}^{\text{1h}}_{\text{ss}} \qty(s_{\parallel} - r_{\parallel}  \big| M), 
\end{align}
and calculate the convolution integral for the satellite-satellite term using the convolution theorem:
\begin{align}
    \lambda_{\text{ss}}(r|M) = \mathscr{F}^{-1}\qty{\big[\tilde{u}_{\text{s}}(k | M)\big]^2}(r),
\end{align}
where $\tilde{u}_{\text{s}}(k | M)$ is the Fourier transform of $u_{\text{s}}(x | M)$ and $\mathscr{F}^{-1}$ denotes the inverse Fourier transform.

\subsection{Two-halo terms}
\label{subsec:xi_R_2h_gcs_emu}

The two-halo term, which accounts for the correlation between galaxies in different haloes, dominates the 2PCF on scales $r \gtrsim 2\text{--}3 \, h^{-1}\mathrm{Mpc}$. 
We decompose redshift-space 2-halo terms using the streaming model:
\begin{align}
    1 + \xi^{\text{S,2h}}_{{\text{ab}}} (s_{\perp}, s_{\parallel}) = \int_{-\infty}^{\infty}\dd{r_{\parallel}} \qty[1 + \xi^{\text{R,2h}}_{{\text{ab}}} (r)] \, \mathcal{P}^{\text{2h}}_{{\text{ab}}}  \qty(v_{\parallel} | \boldsymbol{r}),
\end{align}
where $\mathrm{ab} = \qty{\rm cc, cs, ss}$.
This expression requires models for both the real-space two-halo term, $\xi^{\text{R,2h}}(r)$, and the two-halo pairwise velocity PDF, $\mathcal{P}^{\text{2h}}$.

In the halo model framework, the central-central term can be written as
\begin{align}
    \revision{\xi^{\text{R,2h}}_{\text{cc}}}(r) &= \frac{1}{n_{\text{c}}^2} \int\dd{M_1} \dv{n(M_1)}{M_1} \langle N_{\text{c}}(M_1) \rangle \cdots  \notag \\
    &\phantom{=\frac{1}{n_{}^2}} \int\dd{M_2} \dv{n(M_2)}{M_2} \langle N_{\text{c}}(M_2) \rangle \, \xi^{\text{R}}_{\text{hh}}(r | M_1, M_2). \label{eqn:hm_xiR_2h_cc_expr_ersdv}
\end{align}
The terms involving satellite galaxies will contain convolutions between halo 2PCFs and satellite profiles.
It is simpler to compute them in Fourier space, in which convolutions become products of Fourier modes, then inverse transform back to  configuration space \citep{Cuesta-Lazaro:2023MNRAS.523.3219C},
\begin{align}
    \revision{P^{\text{R,2h}}_{\text{cs}}}(k) &= \frac{1}{n_{\text{c}} n_{\text{s}}} \int \dd{M_1} \frac{\dd{n}(M_1)}{\dd{M_1}} \langle N_{\text{c}} (M_1) \rangle \cdots \notag \\
    &\phantom{= \frac{1}{n_{\text{c}} n_{\text{s}}}} \int \dd{M_2} \frac{\dd{n}(M_2)}{\dd{M_2}} \langle N_{\text{s}}(M_2) \rangle \cdots \notag \\
    &\phantom{= \frac{1}{n_{\text{c}} n_{\text{s}}}} P_{\text{hh}}(k | M_1, M_2) u_{\text{s}}(k | M_2), \label{eqn:hm_PkR_2h_cs_expr_ersdv} \\
    \revision{P^{\text{R,2h}}_{\text{ss}}}(k) &= \frac{1}{n_{\text{s}}^2} \int \dd{M_1} \frac{\dd{n}(M_1)}{\dd{M_1}} \langle N_{\text{s}} (M_1) \rangle \cdots \notag \\
    &\phantom{=\frac{1}{n_{\text{s}}^2}} \int \dd{M_2} \frac{\dd{n}(M_2)}{\dd{M_2}} \langle N_{\text{s}}(M_2) \rangle \cdots \notag \\
    &\phantom{=\frac{1}{n_{\text{s}}^2}} P_{\text{hh}}(k | M_1, M_2) u_{\text{s}}(k | M_1) u_{\text{s}}(k | M_2), \label{eqn:hm_PkR_2h_ss_expr_ersdv}
\end{align}

Now we turn to the line-of-sight pairwise velocity distributions of the 2-halo terms $\mathcal{P}^{\rm 2h}_{{\text{ab}}}(v_{\parallel} | \boldsymbol{r})$.
Following the same spirit of maximally exploiting the statistical symmetry in real-space, we choose to model the joint distribution of radial-transverse pairwise velocities, $\mathcal{P}^{\text{2h}}_{{\text{ab}}}(v_{\rm r}, r_{\rm t} | r \equiv |\boldsymbol{r}|)$, then project it to the line-of-sight direction.
As in previous works \citep{Cuesta-Lazaro:2020MNRAS.498.1175C,Ruan:2022MNRAS.514..440R}, we adopt the skew-T distribution to incorporate the beyond-Gaussian properties of the pairwise velocity distributions, which will be detailed in the following subsection.

\subsection{The skew-T parameterisation of the pairwise velocity distribution}
\label{subsec:pdf_vlos_model_skewt}

An accurate model for the pairwise velocity probability distribution function is essential for applying the streaming model. 
Following recent advancements in the field, we adopt the flexible, four-parameter skewed Student's t-distribution (hereafter ST distribution) proposed by \citet{Azzalini:2009arXiv0911.2342A.skewt,Zu:2013MNRAS.431.3319Z,Cuesta-Lazaro:2020MNRAS.498.1175C}. A key advantage of this model is that its four parameters are uniquely determined by the first four moments of the distribution: the mean, variance, skewness, and kurtosis. This allows the model to capture the non-Gaussian features of the pairwise velocity field robustly.

While the LOS pairwise velocity, $v_{\parallel}$, is the observable quantity, the underlying physics is more naturally described by the joint PDF of the velocity components parallel (radial, $v_{\text{r}}$) and perpendicular (transverse, $v_{\text{t}}$) to the pair separation vector, defined in Equation~\eqref{eqn:vlos_and_vrt_rewfds}. 
Due to statistical isotropy in real space, this joint PDF, $\mathcal{P}(v_{\text{r}}, v_{\text{t}} | r)$, depends only on the separation magnitude $r$. The LOS velocity PDF, $\mathcal{P}(v_{\parallel} | \boldsymbol{r})$, is a projection of this fundamental 2D distribution that depends on the orientation of the pair separation $\boldsymbol{r}$ with respect to the LOS. The relationship between the two, derived from Equation~\eqref{eqn:vlos_and_vrt_rewfds}, is:
\begin{align}
\mathcal{P}(v_{\parallel} | \boldsymbol{r}) = \int\frac{\dd{v_{\text{r}}}}{\sin\theta} \mathcal{P} \qty(v_{\text{r}}, v_{\text{t}}= \frac{ v_{\parallel} - v_{\text{r}}\cos\theta }{ \sin\theta } \bigg| r ).
\end{align}

A PDF can be characterised via its moments. 
For the 2D joint distribution $\mathcal{P}(v_{\text{r}}, v_{\text{t}} | r)$, the raw moments $m_{ij}$ and central moments $c_{ij}$ are defined as:
\begin{equation}
\left\{\begin{aligned}
    m_{ij}(r) & \equiv \iint\dd{v_{\text{r}}} \dd{v_{\text{t}}} (v_{\text{r}})^i (v_\text{t})^j \mathcal{P}(v_{\text{r}}, v_{\text{t}} | r), \\
    c_{ij}(r) &\equiv \iint\!\dd{v_{\text{r}}} \dd{v_{\text{t}}} \!\big[ v_{\text{r}} - m_{10}(r) ]^i \big[ v_{\text{t}} - m_{01}(r) ]^j  \mathcal{P}(v_{\text{r}}, v_{\text{t}} | r).
\end{aligned}\right.
\end{equation}
Due to statistical isotropy, the system is symmetric in the transverse direction, which implies that all moments with an odd power $j$ of the transverse velocity component are zero (e.g. $m_{01}, c_{21}$ and $ c_{03}$).

Similarly, the $n$-th order raw ($m_n$) and central ($c_n$) moments of the 1D LOS velocity distribution are:
\begin{equation}
\left\{\begin{aligned}
    m_n(\boldsymbol{r}) &\equiv \int \dd{v_{\parallel}} (v_{\parallel})^n \,  \mathcal{P}  (v_{\parallel} | r_{\perp}, r_{\parallel}), \\
    c_n(\boldsymbol{r})  &\equiv \int  \dd{v_{\parallel}} \Big[v_{\parallel} - m_1(\boldsymbol{r})\Big]^n \, \mathcal{P}  (v_{\parallel} | r_{\perp}, r_{\parallel}).
\end{aligned}\right.
\end{equation}
The moments of the 1D LOS distribution can be expressed analytically in terms of the moments of the underlying 2D distribution. Up to the fourth order, these projection relations are:
\begin{equation}
\left\{\begin{aligned}
    &m_1 (\boldsymbol{r}) = \mu \, m_{10}(r) \\
    &c_2 (\boldsymbol{r}) = (1 - \mu^2) c_{02}(r) + \mu^2 c_{20}(r) \\
    &c_3 (\boldsymbol{r}) = 3\mu (1 - \mu^2) c_{12}(r) + \mu^3 c_{30}(r) \\
    &c_4 (\boldsymbol{r}) = (1 - \mu^2)^2 c_{04}(r)  + 6 \mu^2 (1 - \mu^2) c_{22}(r) + \mu^4 c_{40}(r),
\end{aligned}\right.\label{eqn:moment_los_project}
\end{equation}
where $\mu \equiv \cos\theta = r_{\parallel}/r$, 
with $\theta$ introduced below Eq.~\eqref{eqn:vlos_and_vrt_rewfds}.  This set of equations allows us to calculate the anisotropic LOS moments from the isotropic radial/transverse moments.

This provides a useful strategy for reducing the sample variances of the velocity moments from simulations. Instead of directly measuring angle-dependent (and thus noisier) LOS moments $c_n(\boldsymbol{r})$, we first measure the simpler isotropic moments of the 2D distribution, which are functions of $r$ only:
\begin{equation}
\left\{
\begin{split}
    &m_{10}(r), \\
    &c_{02}(r), c_{20}(r), \\
    &c_{12}(r), c_{30}(r), \\
    &c_{04}(r), c_{22}(r), c_{40}(r).
\end{split}\right.
\end{equation}
By averaging over all pair orientations at a fixed separation $r$, we can obtain high-precision measurements of these fundamental moments. We then use Equation~\eqref{eqn:moment_los_project} to analytically project these moments and construct the required LOS velocity moments for any orientation $\boldsymbol{r}$.

With the first four LOS moments ($m_1, c_2, c_3, c_4$) in hand, we reconstruct the full velocity PDF using the ST distribution:
\begin{align}
    &\mathcal{P}_{\rmlos}^{\text{ST}} \qty(v_{\parallel} |   v_c (\boldsymbol{r}), w (\boldsymbol{r}), \alpha (\boldsymbol{r}), \nu (\boldsymbol{r})) = \frac{2}{w} t (v_{\parallel} - v_c; \nu) \notag \\
    & \times T\Bigg(
        \alpha \frac{v_{\parallel} - v_c}{w} \bigg[ \frac{\nu + 1}{\nu + \left( ({v_{\parallel} - v_c})/{w}\right)^2} \bigg]^{1/2}; \nu + 1 \Bigg), \label{eq:st_pdf}
\end{align}
where $t (x; \nu)$ is the Student's $t$-distribution with $\nu$ degrees of freedom, and $T(x; \nu + 1)$ is its cumulative distribution function with the number of degrees of freedom $\nu + 1$.
The four parameters of the ST distribution --- location $v_c(\boldsymbol{r})$, scale $w(\boldsymbol{r})$, skewness $\alpha(\boldsymbol{r})$, and tail-weight $\nu(\boldsymbol{r})$ --- are uniquely determined by the four LOS moments we constructed, see Equations~(A1)--(A6) in appendix A of \citet{Cuesta-Lazaro:2020MNRAS.498.1175C}.
\section{Emulators for the model ingredients}
\label{sec:emulators}

The halo streaming model developed here relies on several physical ingredients, such as the halo abundance, clustering, and velocity statistics. 
Directly measuring these quantities from simulations during parameter inference would be computationally prohibitive. 
We therefore construct a suite of emulators that provide fast and accurate predictions across cosmologies, halo mass, and scale. 
Each emulator targets a reduced or physically normalised quantity chosen to minimise dynamic range, improve smoothness, and enforce correct large-scale behaviour. 
Gaussian-process \citep[GP,][]{williams2006gaussian} models are adopted for smooth, low-dimensional functions, while neural networks \citep[NNs,][]{bishop1995neural,Alom:electronics8030292,2021arXiv210611342Z} are used for higher-dimensional quantities with strong scale dependence. 
Table~\ref{tab:emulation_targets} summarises the emulated quantities, their physical definitions, and the surrogate models employed.

\revision{It is worth emphasising the role of this modular design in setting the emulation accuracy achievable from a finite training set of $N_{\text{train}} = 54$ cosmologies. Every emulator targets a \textit{physically normalised residual} rather than the bare statistic: the matter sector is emulated as the signed departure from analytic linear theory, $\Delta\xi_{\text{mm}}(r) = \xi_{\text{mm}}(r) - \xi_{\text{lin}}(r)$ (and the non-linear boost $\alpha(k)$ for $P_{\text{mm}}$), the halo abundance as the ratio $n(>M)/n_{\text{ref}}(>M)$, and the halo clustering as the dimensionless departure from the linear-bias factorisation, $D(r\,|\,M_1,M_2) = \xi_{\text{hh}}/[\,b(M_1)\,b(M_2)\,\xi_{\text{mm}}\,]\rightarrow 1$ at large $r$. These reduced quantities vary far more smoothly with cosmology than the underlying statistics themselves, so that the effective accuracy of the assembled prediction from a fixed number of design nodes is considerably higher than direct emulation of the full statistic would achieve. In practice the leave-one-out errors of these emulators across the design reach the across-realisation noise floor of the training measurements (Sections~\ref{subsec:halo_2pcf} and \ref{subsec:halo_velocity_moments}), i.e.\ at the current simulation volume the size of the training set is not the factor limiting the accuracy. We return in Section~\ref{sec:conclusion} to the accuracy targets of upcoming galaxy-survey analyses and the corresponding extensions of the training suite. We also stress that the matter power spectrum here is an intermediate quantity, used principally to normalise the large-scale halo bias and to convert halo clustering into an effective scale-dependent bias; its percent-level features therefore do not propagate at the same level into the final redshift-space multipoles, where the dominant source of modelling accuracy is the halo emulator rather than the matter emulator.}

\begin{table*}
\centering
\caption{Summary of the emulator targets used in the halo streaming model pipeline. 
For each physical ingredient we list the quantity predicted by the emulator, the reduced form used for training, and the surrogate model adopted. ``PCA'' denotes Principal Component Analysis, ``GP'' denotes Gaussian Process emulators and ``NN'' denotes neural-network emulators.
}
\label{tab:emulation_targets}

\renewcommand{\arraystretch}{1.9}
\begin{tabular}{@{} l l l l @{}}
\toprule
Target Property & Emulated Quantity & Surrogate & Figure \\
\midrule

Matter power spectrum $P_{\text{mm}}(k)$ 
& $\displaystyle \alpha(k) \equiv \frac{P_{\text{mm}}(k)}{P^{\text{lin}}_{\text{mm}}(k)}$
& GP 
& Fig.~\ref{fig:matter_pk_emulator_test} \\

\revision{Matter 2PCF $\xi_{\text{mm}}(r)$}
& \revision{$\displaystyle \Delta\xi(r) \equiv \xi_{\text{mm}}(r) - \xi_{\text{mm}}^{\text{lin}}(r)$}
& \revision{GP}
& \revision{Fig.~\ref{fig:xi_hh_emu}} \\

Halo mass function $\displaystyle \dv{n(M)}{M}$
& $\displaystyle \frac{n(>M)}{n_{\text{ref}}(>M)}$
& \revision{GP (log-PCA)}
& Fig.~\ref{fig:hmf_emulation_wrfsdv} \\


\revision{halo bias $b(M)$}
& --
& \revision{GP}
& \revision{Fig.~\ref{fig:bias_tinker_ext}} \\

\revision{Scale-dependent bias $B(r  | M_1, M_2)$}
& \revision{Equation~\eqref{eqn:scale_dependent_halo_bias_def}}
& \revision{GP (PCA)}
& \revision{Figs.~\ref{fig:xi_hh_emu}, \ref{fig:halo_beta_normalized}} \\

Mean radial pairwise velocity 
& $\displaystyle \frac{m_{10}(r  | M_1,M_2)}{m_{10}^{\text{lin}}(r  | M_1,M_2)}$
& NN 
& Fig.~\ref{fig:halo_vm_m10} \\

\makecell[l]{Pairwise velocity moments \\ (dispersion $n=2$, skewness $n=3$, kurtosis $n=4$)}
& $\displaystyle \frac{c_{(n)}(r  | M_1,M_2)}{\big[m_{10}(r  | M_1,M_2)\big]^n}$
& NN 
& Fig.~\ref{fig:halo_vm_c234} \\

\bottomrule
\end{tabular}
\end{table*}

\subsection{Halo mass functions}
\label{subsec:hmf}

We construct an emulator for the cumulative halo mass function (HMF) using GP regression. This allows us to predict the abundance of haloes as a continuous function of both cosmological parameters and halo mass. 

\subsubsection{Training Data and Input Space}

The training data consist of $N_{\text{train}} = 54$ cosmological models from the simulation suite. 
For each model, we utilise the halo mass function measured in discrete mass bins. 
To ensure the stability of the emulator in the high-mass tail where shot noise is dominant, we restrict the training data to mass scales of $\log_{10}(M [M_\odot/h]) \le 14.8$ at redshift $z = 0.25$.
The cHMFs above the threshold mass bin are modelled separately using analytic Schechter-like fits to suppress high-mass noise, as will be described in Section~\ref{subsubsection:high_mass_hmf_fit}.

\revision{We emulate the cumulative HMF as a curve over halo mass and reduce its dimensionality with a principal-component decomposition (Section~\ref{subsubsec:hmf_gp_impl}), rather than treating halo mass as an additional input of a single global GP. The Gaussian-process inputs are therefore the four cosmological parameters alone,}
\begin{equation}
    \revision{\boldsymbol{x} = \big\{ \Omega_{\text{m0}}, h, S_8, n_{\text{s}} \big\},} \label{eqn:def_x_input_wsfsdsa}
\end{equation}
\revision{with the halo-mass dependence carried by the principal-component basis. We retain only the threshold bins whose mean halo count exceeds ten in every training cosmology, leaving $23$ cumulative thresholds over $12.5 \le \log_{10}(M [h^{-1}M_\odot]) \le 14.7$ with a median relative measurement error of $0.17$ per cent.}

To reduce the dynamic range of the target function and improve the performance of the GP, we emulate the \textit{ratio} of the measured simulation HMF to a theoretical reference model, rather than the number density directly. The target variable $y$ is given by:
\begin{equation}
    y(\boldsymbol{x}) = \frac{n_{\text{sim}}(>M | \boldsymbol{\theta}_{\text{cos}})}{n_{\text{ref}}(>M | \boldsymbol{\theta}_{\text{cos}})}.
\end{equation}
The reference cumulative HMF, $n_{\text{ref}}(>M)$, is calculated using the Sheth-Mo-Tormen mass function \citep{Sheth:2001MNRAS.323....1S} as implemented in the \href{https://hmf.readthedocs.io/en/latest/}{\texttt{hmf}} Python package \citep{Murray:2013A&C.....3...23M.HMFcalc,Murray2021A&C....3600487M.THEHALOMOD}.

\subsubsection{Gaussian Process Implementation}
\label{subsubsec:hmf_gp_impl}

\revision{We reduce the (log) cumulative-HMF ratio $y(\boldsymbol{x})$ curves to a compact representation with a noise-weighted principal-component analysis,}
\begin{equation}
    \revision{\log y(\boldsymbol{x}) \simeq \overline{\log y} + \sum_{k=1}^{K} c_k(\boldsymbol{x})\, \phi_k,}
    \label{eqn:hmf_pca}
\end{equation}
\revision{where the basis $\{\phi_k\}$ is obtained from the mean-subtracted training curves with each output mass bin weighted by the inverse of its measurement error (so that precisely measured bins do not dominate the basis), and each principal-component coefficient $c_k(\boldsymbol{x})$ is modelled by its own Gaussian process over the four-dimensional cosmological input space of Equation~\eqref{eqn:def_x_input_wsfsdsa}. We retain $K=4$ components by an automatic cumulative-variance criterion.}

\revision{Each coefficient process uses an anisotropic radial basis function (squared-exponential) kernel,}
\begin{equation}
    \revision{k(\boldsymbol{x}_i, \boldsymbol{x}_j) = A^2 \exp \left( -\frac{1}{2} \sum_{d=1}^{K} \frac{(x_{i,d} - x_{j,d})^2}{\ell_d^2} \right) + \delta_{ij} \sigma_{\text{n}}^2,}
\end{equation}
\revision{with signal amplitude $A$, characteristic length-scales $\ell_d$ for each of the four cosmological dimensions, and a small diagonal jitter $\sigma_{\text{n}}^2$ added for numerical stability. The hyper-parameters $\{A, \ell_1, \ldots, \ell_4\}$ of each component are determined by maximising the log-marginal likelihood of the training data, using the \href{https://tinygp.readthedocs.io/}{\texttt{tinygp}} package \citep{foreman_mackey_2024_10463641_tinygp}. Because the principal-component decomposition is performed once and each of the $K$ component processes is trained only on the $N_{\text{train}} = 54$ design cosmologies, every Cholesky factorisation involves a covariance matrix of size $N_{\text{train}}$ alone, keeping the GP training cost negligible.}

\revision{Validated against the held-out fiducial cosmology never used in training, this emulator reproduces the cumulative mass function to $0.25$ per cent RMS; its corner-excluded interior leave-one-out error over the design is $0.49$ per cent (median $0.15$ per cent), at the measurement-noise floor of the training data set.}

\subsubsection{The high-mass tail of halo mass functions}
\label{subsubsection:high_mass_hmf_fit}
Direct measurements of the halo mass function from simulations suffer from significant shot noise at the high-mass end where halo number counts are small. To address this problem, we adopt a two-step strategy that prioritises emulation accuracy.

First, we train our emulators exclusively within the mass range $\displaystyle 10^{12.4} \le \qty[M / (h^{-1}M_{\odot})] \le 10^{14.7}$, where the lower limit is dictated by the simulation mass resolution, and the upper limit corresponds to a minimum threshold of $N_{\text{thre}} = 100$ halos in the cumulative mass bin.
By restricting the training data to this well-sampled regime, we ensure high fidelity in the emulation. 
Second, to extend the higher mass range of HMF predictions, we fit a generalised Schechter function to the high-precision emulated HMFs obtained in the first step and extrapolate this parametric form to higher masses.
The parametric form used is:
\begin{equation}
    n(>M) = \phi_\star \left( \frac{M}{M_1} \right)^{-\alpha} \exp\left[ -\left( \frac{M}{M_2} \right)^{\beta} \right],
\label{eq:schechter_gen}
\end{equation}
where $\phi_\star$ is the normalisation constant, $M_1$ and $M_2$ control the characteristic mass scales, $\alpha$ governs the power-law slope, and $\beta$ modifies the exponential cut-off.
To ensure the fit captures the intrinsic shape of the mass function without being biased by resolution effects or shot noise, we restrict the fitting range to mass bins satisfying $N_{\text{thre}} \le N_{\text{halo}} \le 10^3 N_{\text{thre}}$, where $N_{\textrm{halo}}$ is the number of simulation particles in a halo.
The final processed mass function is constructed as a piecewise function.

\begin{figure}
    \centering
    \includegraphics[width=\columnwidth]{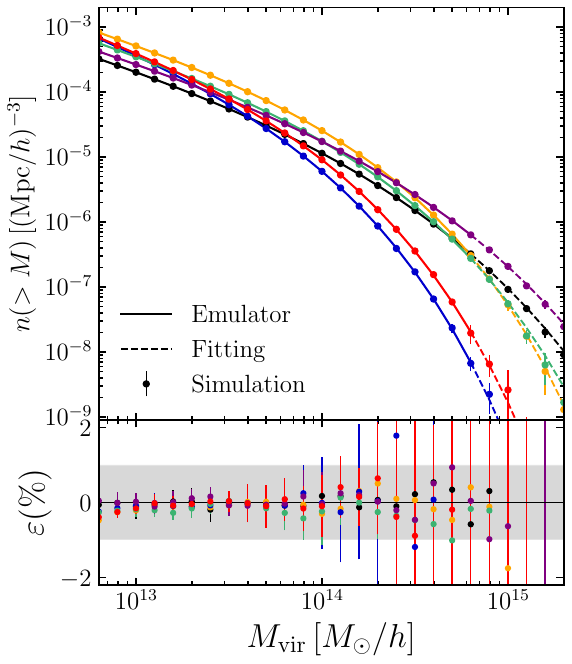}
    \caption{Validation of the cumulative halo mass function (cHMF) emulator against the test data set. \textit{Top panel}: The cumulative halo mass functions for different cosmological models (colour-coded). Points show the ground truth from simulations, solid lines represent the emulator predictions, and dashed lines show the Schechter-fitting extrapolation at the high-mass end. \textit{Bottom panel}: The relative percentage residual between the emulator prediction and the simulation data. \href{https://github.com/chzruan/freyja/blob/main/paper_figs/hmf_gp.py}{\includegraphics[scale=0.019]{figs/freyja.jpg}}}
    \label{fig:hmf_emulation_wrfsdv} 
\end{figure}

Fig.~\ref{fig:hmf_emulation_wrfsdv} illustrates the performance of this enhanced halo mass function emulator across. The upper panel presents the cumulative HMFs, where points represent the measurements directly from $N$-body simulations and solid lines denote the emulator predictions. 
The dashed lines indicate the high-mass regime where the emulator relies on Schechter-function extrapolation. Distinct colours correspond to different cosmologies within the test set. The lower panel quantifies the accuracy of this emulator, showing the relative difference (as a percentage) between the emulated and simulated results.
As shown in the residuals in the lower panel, the emulator achieves sub-percent accuracy over the majority of the mass range.

\subsection{Matter and halo real-space 2-point correlation functions}
\label{subsec:halo_2pcf}

In this subsection, we consider the real-space two-point correlation function of dark matter haloes, $\xi_{\text{hh}}(r | M_1,M_2)$, measured for haloes in mass bins $M_1$ and $M_2$. 
\revision{To maximise emulation accuracy across a wide range of scales, we adopt a modular strategy that separates the modelling of halo clustering from that of halo bias, and that further separates the validated large-scale linear-bias factorisation from a trained small-scale departure. The construction has four ingredients, summarised here and detailed in the following subsections.}

\revision{First, we emulate the nonlinear matter correlation function $\xi_{\text{mm}}(r)$ not from an inverse Fourier transform of the emulated $P_{\text{mm}}(k)$ (which propagates the low-$k$ cosmic-variance noise into the linear regime), but as the signed departure from linear theory, $\Delta\xi(r) \equiv \xi_{\text{mm}}(r) - \xi_{\text{mm}}^{\text{lin}}(r)$, with the linear baseline computed analytically for each cosmology at both training and prediction time (Section~\ref{subsubsec:matter_pk}).}

\revision{Second, on large scales ($r \ge 30\,h^{-1}\mathrm{Mpc}$) we adopt the validated linear-bias factorisation $\xi_{\text{hh}}(r | M_1,M_2) = b(M_1)  b(M_2) \xi_{\text{mm}}(r)$, with bin-averaged effective biases supplied by a peak-height--reparameterised GP emulator of the cumulative bias $b(>M)$ (Sections~\ref{subsec:extrapolation_mass} and \ref{subsubsec:xi_hh}).}

\revision{Third, on nonlinear and quasi-linear scales ($2 \le r < 30\,h^{-1}\mathrm{Mpc}$) we emulate the dimensionless departure from this factorisation,}
\begin{align}
    \revision{D(r | {>}M_1, {>}M_2)} &\revision{{}\equiv \frac{\xi_{\text{hh}} (r | {>}M_1, {>}M_2)}{b({>}M_1)\, b({>}M_2)\, \xi_{\text{mm}} (r)} \;\longrightarrow\; 1,}
    \label{eqn:scale_dependent_halo_bias_def}
\end{align}
\revision{at large $r$, defined at the level of \textit{cumulative mass thresholds} (Section~\ref{subsubsec:xi_hh}). Cumulative thresholds aggregate the halo counts, so this target is far less noisy than the differential-bin scale-dependent bias $\mathcal{B}(r | M_1, M_2) \equiv \xi_{\text{hh}}/[b(M_1) b(M_2) \xi_{\text{mm}}]$ used in the original construction; halo mass enters as an output dimension rather than as an emulator input.}

\revision{Fourth, predictions for \textit{arbitrary} differential mass bins are obtained from the threshold-level prediction through an exact count-weighted mixed-difference identity, with every number density supplied by the cumulative HMF emulator (Section~\ref{subsubsec:xi_hh}); the differencing is performed losslessly at the pair-count level, so the binned $\xi_{\text{hh}}$ entering the halo-model mass integrals is exactly consistent with the mass function that weights them. This formulation isolates the mass-dependent, nonlinear halo physics into a smoother quantity, reducing the dynamic range and noise level of the emulation target, following the idea of emulating the HMF ratio presented above.}

\revision{The underlying simulation measurements of $\xi_{\text{hh}}$ are made in differential halo-mass bins of width $0.1$~dex spanning $12.2 \le \log_{10}(M [h^{-1}M_\odot]) \le 15.0$ (29 bins), identical to the binning of the pairwise velocity moments (Section~\ref{subsec:halo_velocity_moments}). For the original differential-bin emulation we retained the well-sampled subset $12.4 \le \log_{10}M \le 14.0$, with the Tinker-calibrated bias extrapolation (Section~\ref{subsec:extrapolation_mass}) covering higher masses. In the rebuilt pipeline the small-scale departure $D$ is instead defined on cumulative mass thresholds spaced by $0.05$~dex, and the exact threshold-to-bin conversion (Section~\ref{subsubsec:xi_hh}) returns predictions for arbitrary mass-bin edges at evaluation time, so the prediction binning is no longer tied to the training binning.}

\subsubsection{Matter power spectrum}
\label{subsubsec:matter_pk}

An accurate description of the nonlinear matter power spectrum $P_{\text{mm}}(k)$, and its Fourier counterpart, $\xi_{\text{mm}}(r)$, are needed to model halo clustering on large scales, where the matter field provides the baseline signal entering the halo bias and streaming components of the model. Rather than directly emulating the full nonlinear matter power spectra measured from simulations, we construct emulators for the nonlinear boost factor
\begin{align}
    \alpha(k) \equiv \frac{P_{\text{mm}}(k)}{P_{\text{mm}}^{\text{lin}}(k)} ,
\end{align}
where $P_{\text{mm}}^{\text{lin}}(k)$ denotes the linear matter power spectrum, which is computed using the Boltzmann solver \href{https://camb.readthedocs.io/en/latest/}{\texttt{CAMB}} \citep{Lewis:1999bs.camb} for each cosmological model in the training and prediction stages.

This choice provides several advantages. First, the linear power spectrum already captures, to some degree, the cosmology dependence through well-understood linear perturbation theory, reducing the dynamic range of the training data. 
Second, the ratio $\alpha(k)$ approaches unity on large scales ($k \rightarrow 0$), which stabilises the learning problem and significantly improves accuracy in the low-$k$ regime where small absolute errors in $P_{\text{mm}}(k)$ would otherwise translate into large relative deviations. 
The emulator predicts $\alpha(k)$, which is then combined with the corresponding linear spectrum to reconstruct the nonlinear matter power spectrum.

\begin{figure}
    \centering
    \includegraphics[width=\columnwidth]{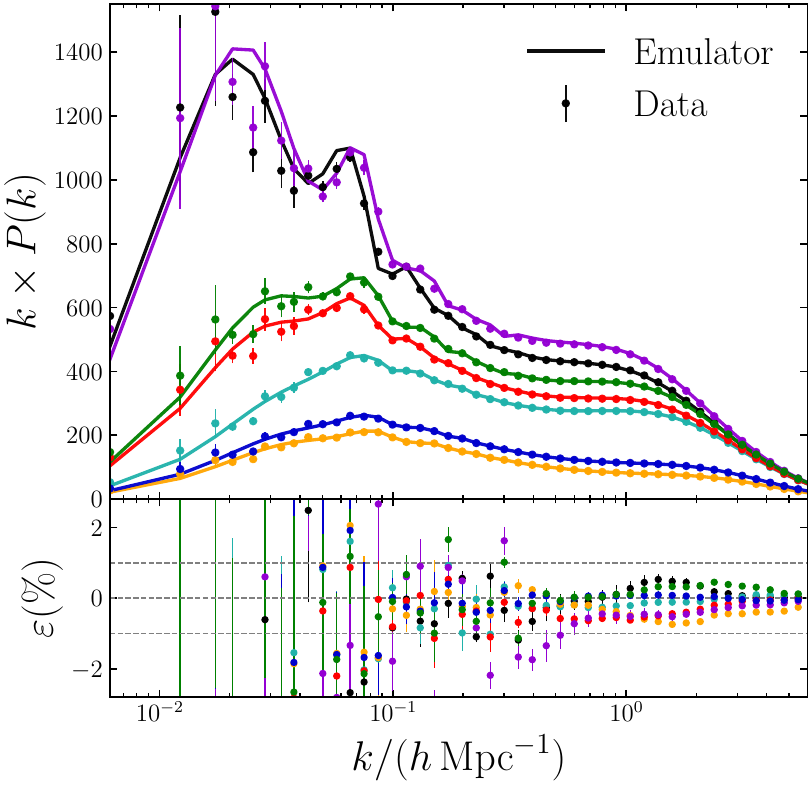}
    \caption{
        Validation of the matter power spectrum emulator. 
        \textit{Top panel:} comparison between emulator predictions (solid lines) and simulation measurements (points) for the cosmologies in the test data set. 
        Colours encode different cosmological models. 
        \textit{Bottom panel:} relative difference between emulator and simulation results. 
        The emulator reproduces the nonlinear matter power spectrum with an accuracy of approximately $1\%$ across the scale range used in the analysis.
        \href{https://github.com/chzruan/freyja/blob/main/paper_figs/matter_pk_emulator.py}{\includegraphics[scale=0.019]{figs/freyja.jpg}}
    }
        \label{fig:matter_pk_emulator_test}
\end{figure}

Fig.~\ref{fig:matter_pk_emulator_test} presents validation results on an independent test set not used during training. The upper panel compares emulator predictions (solid curves) with measurements from simulations (points), demonstrating excellent agreement across the entire $k$ range considered. \revision{The lower panel shows the relative residuals. The interior leave-one-out error is $0.41$ per cent RMS over $0.1 \le k/(h\,\mathrm{Mpc}^{-1}) \le 6$, and the out-of-sample error on the held-out fiducial cosmology is $0.89$ per cent RMS in this window (dominated by the cosmic variance of the single five-box fiducial at low $k$; $0.2$--$0.6$ per cent at $k>0.1$ against the 100-box fiducial); the full set of per-emulator figures of merit is collected in Table~\ref{tab:emulator_accuracy}. At $k \lesssim 0.03\,h\,\mathrm{Mpc}^{-1}$ the boost is smoothly anchored to linear theory, removing the low-$k$ cosmic-variance noise of the finite-volume measurements.} This level of precision is sufficient for propagating matter clustering uncertainties into the halo-model predictions without introducing a dominant systematic error.

While baryon acoustic oscillation (BAO) features are clearly present in the matter power spectra, our primary objective here is accurate modelling of nonlinear galaxy clustering rather than precision reconstruction of the BAO signal itself. 
Consequently, we do not introduce additional preprocessing steps specifically designed to optimise the emulation of oscillatory features. 
\revision{In configuration space, the BAO feature and the large-scale zero crossing are instead carried by the analytic linear-theory baseline of the $\xi_{\text{mm}}(r)$ emulator (Section~\ref{subsec:halo_2pcf}): because we emulate only the smooth departure $\Delta\xi = \xi_{\text{mm}} - \xi_{\text{mm}}^{\text{lin}}$, the Gaussian process is never asked to learn the oscillatory or zero-crossing features, and the noise-normalised residual through the BAO band is $\chi_{\text{rms}} \simeq 0.46$, i.e.\ at the measurement noise floor. A dedicated BAO analysis would treat the feature more carefully (e.g.\ with a BAO-template fit), which is straightforward to add within the modular pipeline but is beyond the scope of this paper.}

\revision{We construct our own matter and HMF emulators, rather than adopting existing public ones, primarily to ensure full self-consistency in the cosmology dependence: deriving every ingredient from the same simulation suite means that cross-emulator residuals (e.g.\ between $P_{\text{mm}}(k)$ and $\xi_{\text{hh}}/\xi_{\text{mm}}$) cancel rather than add, and the cosmology grids match exactly within a single likelihood. However, the modular design of the pipeline allows any of the matter or HMF emulators to be replaced by an external one, such as \texttt{EuclidEmulator2} \citep{EuclidEmulator2:2021MNRAS.505.2840E} for the nonlinear matter power spectrum and dedicated halo-mass-function emulators \citep[e.g.][]{McClintock:2019ApJ...872...53M.AemulusHMF,Bocquet:2020ApJ...901....5B.MTHMF,2024A&A...691A.323S.e-MANTIS,Chen:2025SCPMA..6809513C.HMFEmu}.}

\subsubsection{Large-scale linear bias $b(M)$}

On sufficiently large scales ($r \ge 40\,h^{-1}\mathrm{Mpc}$), halo clustering is expected to trace the underlying matter distribution in a linear, scale-independent manner. 
In this regime, the halo-halo correlation function can be approximated as
\begin{equation}
    \xi_{\text{hh}}(r | M_1, M_2) \approx b_{1}(M_1) b_2  (M_2) \, \xi_{\text{mm}}(r),
    \label{eqn:linear_bias}
\end{equation}
where $b(M)$ denotes the linear halo bias. 
We estimate $b(M)$ directly from the simulations by fitting the ratio of measured correlation functions,
\begin{equation}
    b(M) = \left\langle \qty[\frac{\xi_{\text{hh}} (r | M, M)}{\xi_{\text{mm}}(r)]}]^{1/2} \right\rangle_{r},
\end{equation}
where the average is taken over the separation range $40 \le r/(h^{-1}\mathrm{Mpc}) \le 80$. 
The average weights are chosen to be inversely proportional to the propagated variance of the ratio across the five simulation realisations of each cosmological model. This ensures that scales with higher signal-to-noise ratios dominate the fit.
This range is selected to exclude nonlinear halo exclusion effects at smaller separations while limiting the impact of sample variance at very large scales in finite-volume simulations.

\subsubsection{Mass extrapolation}
\label{subsec:extrapolation_mass}

A major challenge in emulating the mass dependence of halo clustering statistics arises from the scarcity of high-mass haloes. 
In the \textsc{DEGRACE}-pilot simulations, shot noise becomes dominant for halo mass $M \gtrsim 10^{14}\,h^{-1}M_\odot$, rendering direct emulation of $\xi_{\text{hh}}(r | M_1,M_2)$ reliable only in a narrow mass range. 
To enable robust predictions for these rare objects, we implement a physically motivated extrapolation scheme based on the linear halo bias.

To extend the bias model into the cluster-mass regime ($M = 10^{14} \text{--} 10^{15}\,h^{-1}M_\odot$), where direct measurements are already noise dominated, we adopt the functional form proposed by \citet{Tinker:2010ApJ...724..878T}, which expresses the bias as a function of the peak height $\nu$,
\begin{align}
    b(\nu) = 1 - A \frac{\nu^a}{\nu^a + \delta_{\text{c}}^a} + B \nu^b + C \nu^c, \label{eqn:tinker_bias_esrffffsdv}
\end{align}
where $\delta_{\text{c}} = 1.686$ is the critical overdensity for spherical collapse. 
The peak height is defined as $\nu \equiv \delta_{\text{c}}/\sigma(M)$, with $\sigma(M)$ denoting the root-mean-square fluctuation of the linear density field smoothed on the Lagrangian scale $R(M) = [3M/(4\pi\bar{\rho}_{\text{m}})]^{1/3}$.

We calibrate this model by fitting Equation~\eqref{eqn:tinker_bias_esrffffsdv} to the simulated bias measurements in the well-sampled mass range $10^{12.4} < M / (h^{-1}M_\odot) < 10^{13.8}$. To avoid unphysical behaviour in the high-mass extrapolation, we adopt a constrained fitting strategy in which the shape parameters controlling the asymptotic scaling at large $\nu$ are fixed close to the reference values of \citet{Tinker:2010ApJ...724..878T} ($a=0.132$, $b=1.5$, $c=2.4$), while allowing only the normalisation coefficients $(A,B,C)$ to vary. 
This ensures that the extrapolated bias retains the expected $\nu^2$ scaling for rare density peaks \citep{Sheth:1999MNRAS.308..119S.PBS}, while still remaining flexible enough to accommodate the halo definition and cosmology variations in the simulation suite adopted here. The resulting calibrated model is then used to predict $b(M)$ for all haloes with $M \ge 10^{13.8}\,h^{-1}M_\odot$.

\begin{figure}
    \centering
    \includegraphics[width=\columnwidth]{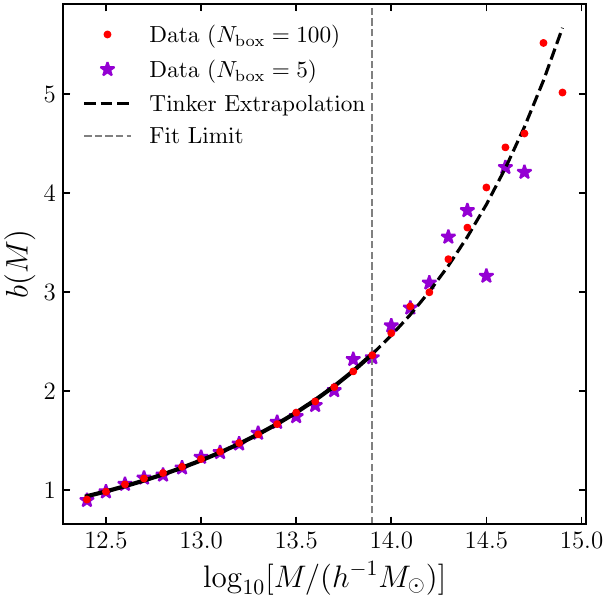}
    \caption{
        Linear halo bias as a function of halo mass at $z=0.25$ for the fiducial \textit{Planck} cosmology \citep{Planck15Parameters:2016A&A...594A..13P}. 
        The purple stars represent measurements from a subset of five realisations ($N_{\text{box}}=5$), mimicking the noise level of the training data set. 
        The red points show the ground truth derived from the full suite of $100$ realisations.
        The black curve follows the \citet{Tinker:2010ApJ...724..878T} bias model: the solid section marks the range used to fit the normalisation parameters, while the dashed section to the right of the vertical line represents the extrapolation to higher masses
        The vertical grey dashed line indicates the upper mass limit of the fitting region. 
        The extrapolation demonstrates excellent agreement with the high-mass validation data.~\href{https://github.com/chzruan/freyja/blob/main/paper_figs/bias_tinker_extrapolation.py}{\includegraphics[scale=0.019]{figs/freyja.jpg}}
        }
    \label{fig:bias_tinker_ext}
\end{figure}

Fig.~\ref{fig:bias_tinker_ext} illustrates the behaviour of the linear halo bias as a function of halo mass at $z=0.25$ for the fiducial \textit{Planck} cosmology \citep{Planck15Parameters:2016A&A...594A..13P}. 
Measurements obtained from a reduced set of five simulation realisations are shown in purple stars as representative of the noise level in the emulator training data.
In contrast, the bias measured from the full ensemble of $100$ realisations (red points) provides a high-precision reference against which the modelling can be assessed. 
The black curve shows the bias predicted by the \citet{Tinker:2010ApJ...724..878T} fitting formula, with the solid segment indicating the mass range used to calibrate the parameters. 
The vertical grey dashed line marks the upper boundary of this fitting region, beyond which the model is extrapolated (black dashed curve). 
The extrapolated bias remains in excellent agreement with the high-mass validation measurements, demonstrating that the parametric model, when anchored on the well-measured low-mass regime, provides a reliable mass extension of halo bias and 2PCF in the sparse cluster-mass tail.

\subsubsection{The scale-dependent bias and the full halo two-point correlation function}
\label{subsubsec:xi_hh}

\begin{figure}
    \centering
    \includegraphics[width=\columnwidth]{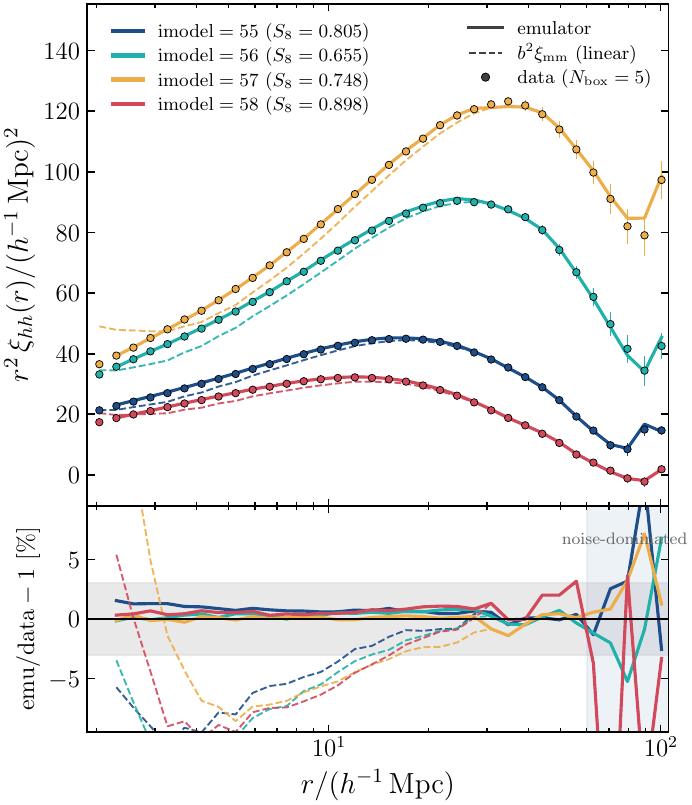}
    \caption{
    \revision{Halo correlation function
      $\xi_{\text{hh}}(r)$ for haloes above $\log_{10}[M/(h^{-1}M_\odot)] = 12.5$ at $z=0.25$ in $\Lambda$CDM, for four test cosmologies 55--58. Points show the measured correlation function and solid curves the emulator prediction.
      Dashed curves corresponds to the large-scale linear bias approximation. The lower panel shows the fractional residual, emulator/data $-1$; the grey band marks $\pm3$ per cent and the shaded region $r>60\,h^{-1}\mathrm{Mpc}$ is dominated by the measurement noise. The emulator reproduces the measured $\xi_{\text{hh}}$ to $0.5$--$0.9$ per cent RMS over $r<60\,h^{-1}\mathrm{Mpc}$ across the test data set.}  ~\href{https://github.com/chzruan/freyja/blob/main/paper_figs/plot_xi_hh_massrange.py}{\includegraphics[scale=0.019]{figs/freyja.jpg}}
    }
    \label{fig:xi_hh_emu}
\end{figure}

Fig.~\ref{fig:xi_hh_emu} presents a representative validation of the halo 2-point correlation function emulator using a cosmological model drawn from the independent test data set. The comparison demonstrates that the emulator accurately reproduces the real-space halo clustering measured in the $N$-body simulations across a wide range of separations. 

On small and intermediate scales ($2 \le r < 30\,h^{-1}\mathrm{Mpc}$), the prediction is governed by the threshold-level departure emulator $D(r | {>}M_1, {>}M_2)$ defined in Equation~\eqref{eqn:scale_dependent_halo_bias_def}, which captures nonlinear halo exclusion and scale-dependent clustering effects. \revision{This surface is emulated on $465$ ordered cumulative-threshold pairs (mass thresholds spaced by $0.05$~dex up to $10^{14}\,h^{-1}M_\odot$) via a noise-weighted linear Principal Component Analysis (PCA) with $K=8$ components, each carrying its own Gaussian process; columns within the halo-exclusion wedge, $r < 1.5\,[R_{\text{vir}}(M_1)+R_{\text{vir}}(M_2)]$, are masked in training and propagated as not-a-number in prediction, since on those scales the signal is dominated by exclusion physics that a smooth emulator should not learn.} Toward large separations ($r \ge 30\,h^{-1}\mathrm{Mpc}$), the model transitions smoothly to the \revision{validated} linear-bias \revision{factorisation, with $D \equiv 1$}. This stitching procedure enforces physical consistency in the linear regime while preserving the flexibility of the nonlinear emulator on smaller scales. 

\revision{Predictions for arbitrary differential mass bins are produced at evaluation time from the threshold-level surface through an exact count-weighted mixed-difference identity. Writing 
\begin{align}
    F(r;{>}M_1,{>}M_2) \equiv n({>}M_1)\,n({>}M_2)\,[1+\xi_{\text{hh}}(r|{>}M_1,{>}M_2)]
\end{align}
for the (volume-normalised) threshold pair count, the binned correlation function between mass bins $[M_a,M_b)$ and $[M_c,M_d)$ follows losslessly from the mixed difference}
\begin{align}
  \revision{1+\xi_{\text{hh}}\bigl(r | [M_a,M_b),[M_c,M_d)\bigr) = \frac{F_{ac}-F_{bc}-F_{ad}+F_{bd}}{\Delta n_{ab}\,\Delta n_{cd}},}
  \label{eqn:xihh_mixdiff}
\end{align}
\revision{with $\Delta n_{ab} = n({>}M_a)-n({>}M_b)$ and every number density $n({>}M)$ supplied by the cumulative HMF emulator of Section~\ref{subsec:hmf}. Because the differencing is performed at the count level, the binned $\xi_{\text{hh}}$ entering the halo-model mass integrals is exact and consistent with the mass function that weights those integrals by construction. 
}

\begin{figure*}
    \centering
    \includegraphics[height=0.45\textwidth]{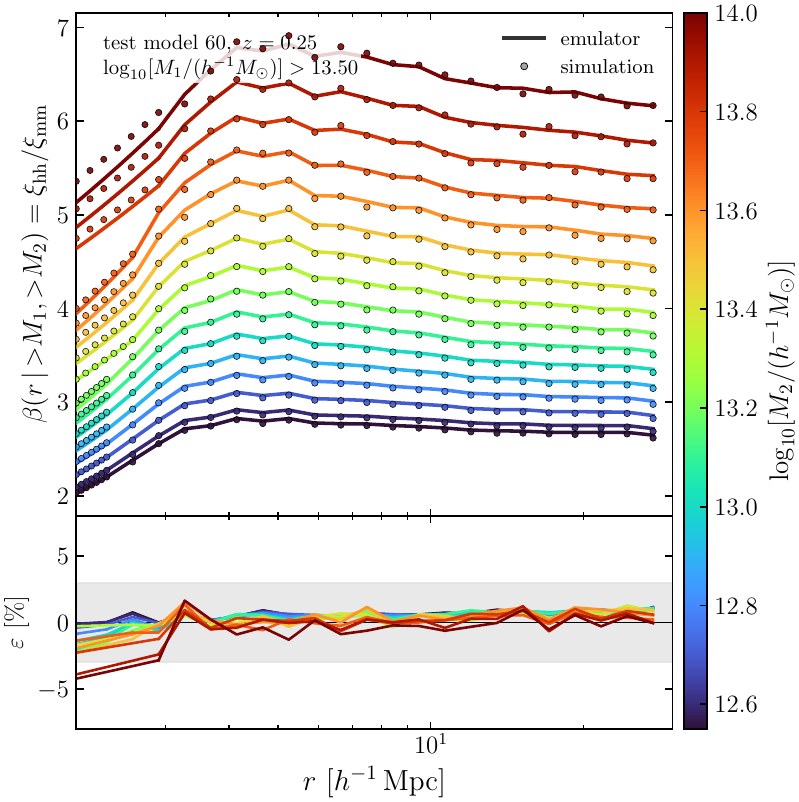}\quad
    \includegraphics[height=0.45\textwidth]{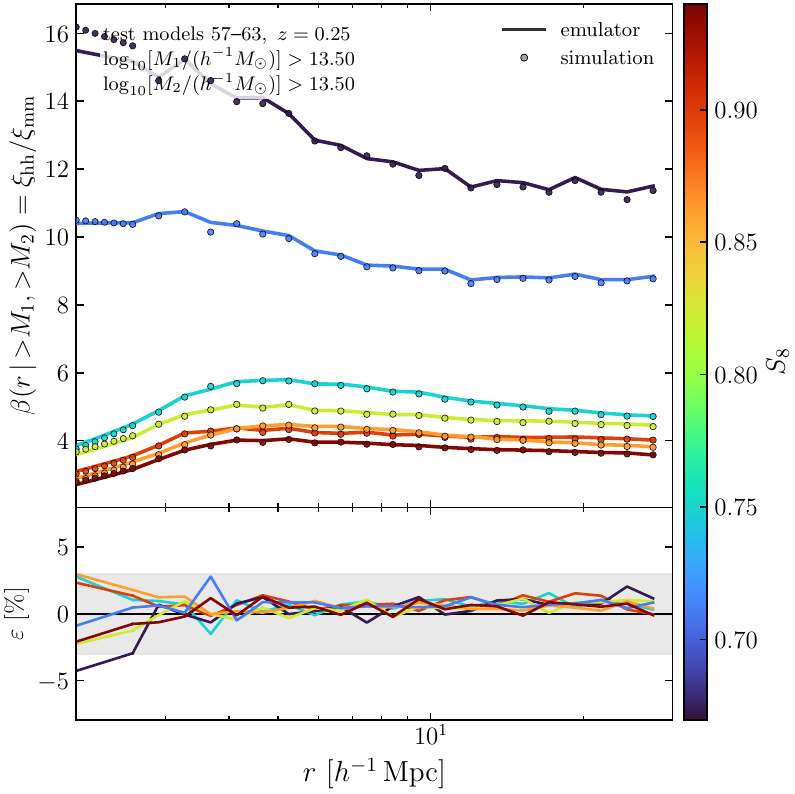}
    \caption{
        Emulator predictions for the scale-dependent halo bias. 
        \textit{Left panel:} \revision{fixed cosmology (test model~60) and} $M_1 = 10^{13.5}\,h^{-1}M_{\odot}$ with varying $M_2$, as indicated by the colour bar. 
        \textit{Right panel:} fixed halo masses ($M_1=M_2=10^{13.5}\,h^{-1}M_{\odot}$) with varying cosmologies \revision{(the seven held-out test models~57--63)}. The lower panel shows the residuals between the emulator and the direct measurement. 
        \revision{The emulator captures smooth variations with both halo mass and cosmology. Because the emulator is trained on cumulative mass thresholds rather than differential bins, the training data are far less noisy; arbitrary differential mass bins are then recovered losslessly through the exact count-weighted threshold-to-bin conversion of Equation~\eqref{eqn:xihh_mixdiff}.}
        ~\href{https://github.com/chzruan/freyja/blob/main/paper_figs/halo_beta_normalized_plots.py}{\includegraphics[scale=0.019]{figs/freyja.jpg}}
    }
    \label{fig:halo_beta_normalized}
\end{figure*}

\revision{Fig.~\ref{fig:halo_beta_normalized} illustrates the behaviour of the emulator, showing the departure surface $\xi_{\text{hh}}(r | M_1, M_2) / \xi_{\text{mm}}(r)$. To visualise its parameter dependence, the left panel shows results in a fixed cosmology from the test set and fixed mass $M_1 = 10^{13.5}\,h^{-1}M_{\odot}$ while varying $M_2$. The right panel fixes both halo masses at $M_1=M_2=10^{13.5}\,h^{-1}M_{\odot}$ and varies the cosmological model across the test sample. Defining the emulation target on cumulative thresholds rather than on differential mass bins reduces the measurement noise of the training data (median $0.8$ per cent), so the subset leave-one-out error of the $D$ surface is $0.016$ in absolute $D$ units.}

\revision{The decisive end-to-end test reconstructs the directly measured differential-bin $\xi_{\text{hh}}$ from the full pipeline across ten test cosmologies. As summarised in Table~\ref{tab:emulator_accuracy} and shown in Fig.~\ref{fig:xi_hh_emu}, the rebuilt route reproduces the binned $\xi_{\text{hh}}$ at $1.0$--$1.4$ per cent median accuracy in every scale band below $30\,h^{-1}\mathrm{Mpc}$, compared with $3.7$--$9.1$ per cent for the pure factorisation on the same scales; on the held-out fiducial cosmology individual mass bins are reproduced at $1.4$--$3.7$ per cent RMS over $r \le 30 \,h^{-1}\mathrm{Mpc}$. A single full-scale call over $r \in [2,100]\,h^{-1}\mathrm{Mpc}$ matches the measurements at the data noise floor everywhere, with a noise-normalised residual of median $|\chi| \simeq 0.5$--$1$ in every band, so the quoted percentages are upper bounds set by the measurement noise rather than by the emulator. Because the binned $\xi_{\text{hh}}$ is exactly consistent with the HMF that weights it (Equation~\ref{eqn:xihh_mixdiff}), this accuracy is achieved without relying on the cancellation of residuals inside the halo-model mass integrals.}

\begin{table*}
\centering
\caption{\revision{Summary of emulator accuracy for the model ingredients. ``Interior LOO'' is the corner-excluded interior leave-one-out RMS fractional error over the design; ``OOS'' is the RMS fractional residual on the held-out fiducial cosmology, never used in training. The binned $\xi_{\text{hh}}$ row reports the median $|\mathrm{model}/\mathrm{measured}-1|$ of the reconstructed differential-bin correlation function over $780$ (cosmology, bin-pair) combinations, compared with the pure linear-bias factorisation on the same scales; over the full range its noise-normalised residual is median $|\chi| \simeq 0.5$--$1$, i.e.\ at the measurement noise floor.}}
\label{tab:emulator_accuracy}
\renewcommand{\arraystretch}{1.4}
\begin{tabular}{@{} l l l l @{}}
\toprule
\revision{Emulated quantity} & \revision{Surrogate} & \revision{Interior LOO} & \revision{Out-of-sample (fiducial)} \\
\midrule
\revision{Cumulative HMF $n({>}M)$} & \revision{log-PCA + GP} & \revision{$0.49\%$ (median $0.15\%$, at SEM floor)} & \revision{$0.25\%$} \\
\revision{Matter power spectrum $P_{\text{mm}}(k)$} & \revision{$P/P_{\text{lin}}$ boost, low-$k\!\to$linear} & \revision{$0.41\%$ over $k \in [0.1,6]$} & \revision{$0.89\%$ (window)} \\
\revision{Matter 2PCF $\xi_{\text{mm}}(r)$} & \revision{noise-weighted PCA + GP} & \revision{$0.37\%$ over $[30,60]\,h^{-1}\mathrm{Mpc}$} & \revision{$0.28\%$ (masked RMS)} \\
\revision{Cumulative bias $b({>}M)$} & \revision{peak-height GP + Tinker tail} & \revision{$\simeq 0.5\%$ (at noise floor)} & \revision{---} \\
\revision{Departure surface $D(r|{>}M_1,{>}M_2)$} & \revision{noise-weighted PCA + GP} & \revision{$0.016$ (absolute $D$, subset LOO)} & \revision{---} \\
\revision{Binned $\xi_{\text{hh}}$ (rebuilt vs.\ factorised)} & \revision{threshold-to-bin (Eq.~\ref{eqn:xihh_mixdiff})} & \revision{$1.0$--$1.4\%$ vs.\ $3.7$--$9.1\%$ ($r<30$)} & \revision{$1.4$--$3.7\%$ over $[2.2,28]$} \\
\bottomrule
\end{tabular}
\end{table*}

\subsection{Halo pairwise velocity moments}
\label{subsec:halo_velocity_moments}

We construct emulators for halo pairwise velocity moments up to fourth order,
\begin{align}
    \revision{\{ m_{10}^{\text{hh}},\, c_{20}^{\text{hh}},\, c_{02}^{\text{hh}},\, c_{30}^{\text{hh}},\, c_{12}^{\text{hh}},\, c_{40}^{\text{hh}},\, c_{22}^{\text{hh}},\, c_{04}^{\text{hh}} \}}(r | M_1, M_2),
\end{align}
which fully specify the non-Gaussian pairwise velocity statistics entering the halo streaming model. 
The measurements are obtained from $N$-body simulations on differential halo mass grids and averaged over five realisations to reduce sample variance. 
As velocity moments exhibit large dynamic range and strong cosmology dependence, we adopt a physics-informed rescaling before emulation.

On large scales, the mean radial pairwise velocity follows from linear perturbation theory and pair conservation. 
Assuming unbiased halo velocities, the linear-theory prediction for two halo populations with masses $(M_1, M_2)$ is
\begin{align}
    m_{10}^{\text{hh,lin}}(r | M_1,M_2)
    =
    -\frac{2}{3}\, a H f\, r
    \left(\frac{b(M_1)+b(M_2)}{2}\right)
    \bar{\xi}_{\text{mm}}(r),
\end{align}
where $f=d\ln D/d\ln a$ is the linear growth rate and $\bar{\xi}_{\text{mm}}(r)$ is the matter correlation function averaged over a sphere of radius $r$. 
In practice, $\bar{\xi}_{\text{mm}}$ is computed from the linear matter correlation function predicted by the matter $\alpha$ emulator.
This construction provides a physically motivated large-scale anchor for the velocity field. 
Rather than emulating $m_{10}$ directly, we define a reduced quantity
\begin{align}
    \omega(r | M_1, M_2)
    \equiv
    \frac{\revision{m_{10}^{\text{hh}}}(r | M_1,M_2)}
         {\revision{m_{10}^{\text{hh,lin}}} (r | M_1,M_2)},
\end{align}
which approaches unity on large scales and varies smoothly with cosmology and halo mass. 
This physics-informed normalisation substantially improves emulator stability and extrapolation behaviour.

Higher-order moments span several orders of magnitude and may change sign across scales. 
To stabilise the learning problem, all velocity moments are expressed in terms of reduced variables scaled by powers of the first-order moment $m_{10}$,
\begin{align}
    \qty(\frac{c_{20}}{m_{10}^2})^{1/2}, \left(\frac{c_{30}}{m_{10}^3}\right)^{1/3}, \left(\frac{c_{40}}{m_{10}^4}\right)^{1/4},
\end{align}
with analogous scalings applied to other higher-order moments.

The emulation is performed using a neural-network model following the same modular design as the halo bias emulator. 
\revision{Specifically, both the scale-dependent-bias and the velocity-moment emulators are fully-connected feed-forward networks (multilayer perceptrons) with two hidden layers of $256$ units each and GELU activations, taking as inputs $(\Omega_{\text{m0}}, h, S_8, n_{\text{s}}, \log_{10} M_1, \log_{10} M_2, \log r)$. They are trained by minimising an inverse-variance-weighted mean-squared-error loss, with the per-point weights set to $1/\sigma_i^2$ from the across-realisation scatter, using the AdamW optimiser (initial learning rate $10^{-3}$, weight decay $10^{-4}$) with a reduce-on-plateau schedule that halves the learning rate after $10$ stagnant epochs, a batch size of $128$, no dropout, and early stopping on the validation loss with a patience of $50$ epochs.} 
\revision{The input and target variables are whitened (subtracting the mean and dividing by the standard deviation) using normalisation statistics computed on the training data set, which are then applied unchanged to the validation and test sets; in this way no information from the validation or test data enters the training procedure, which would otherwise yield optimistic estimates of the generalisation accuracy.} 
After prediction, the inverse transformation is applied to recover the physical velocity moments required by the streaming model.

\begin{figure*}
    \centering
    \includegraphics[height=0.45\textwidth]{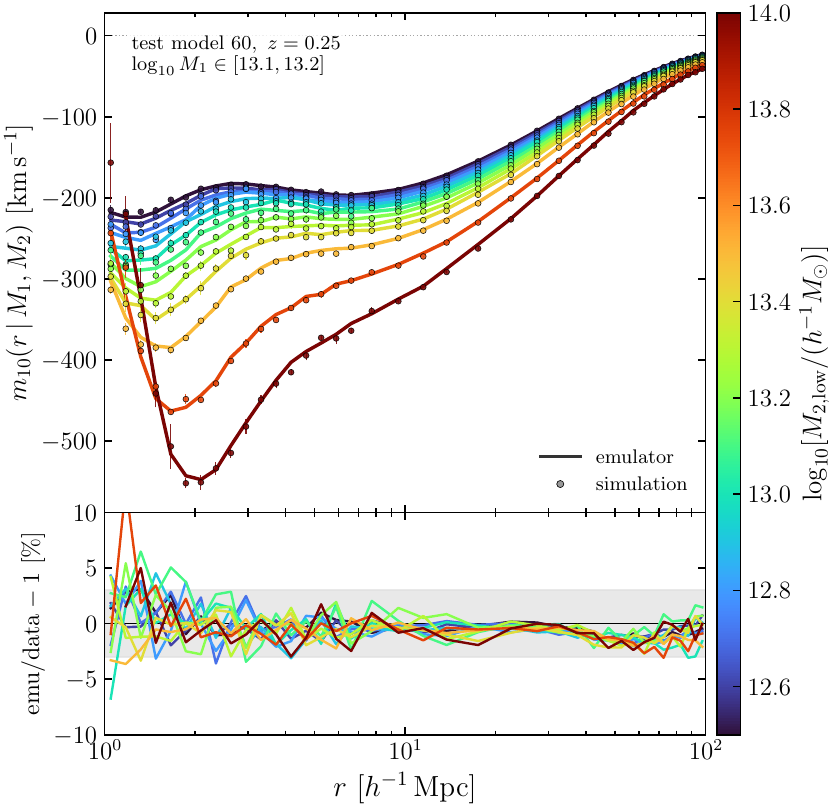}\quad %
    \includegraphics[height=0.45\textwidth]{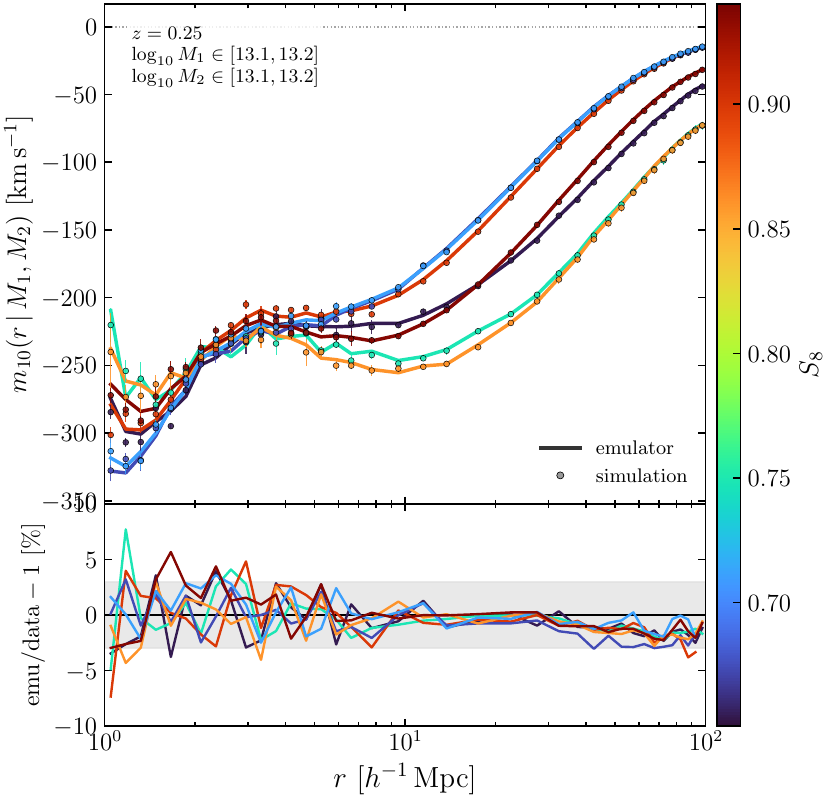}
    \caption{
        Emulator predictions for the mean halo pairwise velocity moment $m_{10}(r | M_1,M_2)$.
        \textit{Left panel:} \revision{fixed cosmology (test model~60) with} $M_1 = 10^{13.1}\,h^{-1}M_{\odot}$ and varying $M_2$ as indicated by the colour bar.
        \textit{Right panel:} fixed halo masses ($M_1=M_2=10^{13.1}\,h^{-1}M_{\odot}$) with varying cosmologies \revision{(the seven held-out test models~57--63), with the line colour keyed to $S_8$ via the colour bar}.
        The emulator reproduces smooth mass and cosmology dependence despite noise in differential measurements. \revision{The largest residuals correlate weakly with distance from the centre of the training prior in $S_8$, as expected for an interpolating emulator.}
        ~\href{https://github.com/chzruan/freyja/blob/main/paper_figs/halo_vm_plots.py}{\includegraphics[scale=0.019]{figs/freyja.jpg}}
    }
    \label{fig:halo_vm_m10}
\end{figure*}

Fig.~\ref{fig:halo_vm_m10} shows emulator predictions for the mean pairwise velocity. 
The emulator captures smooth variations with halo mass and cosmology despite the intrinsic noise present in differential mass-bin measurements. 
As in the halo clustering emulator, residual errors fluctuate around zero and partially cancel once integrated within halo-model expressions, yielding stable predictions for galaxy clustering observables.

\begin{figure*}
    \centering
    \includegraphics[width=0.33\textwidth]{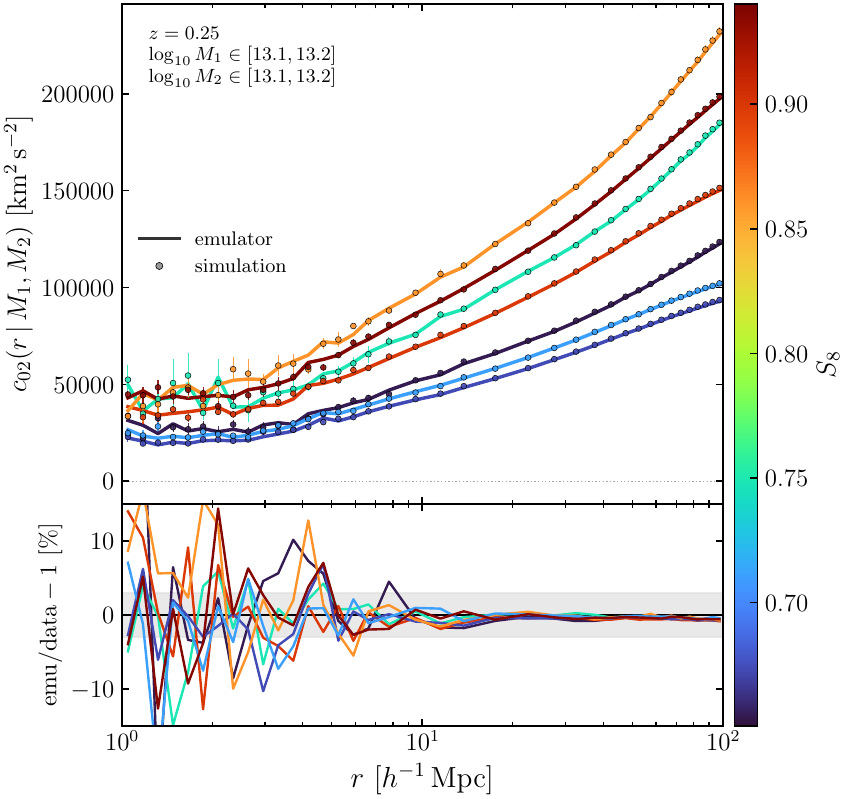}%
    \includegraphics[width=0.33\textwidth]{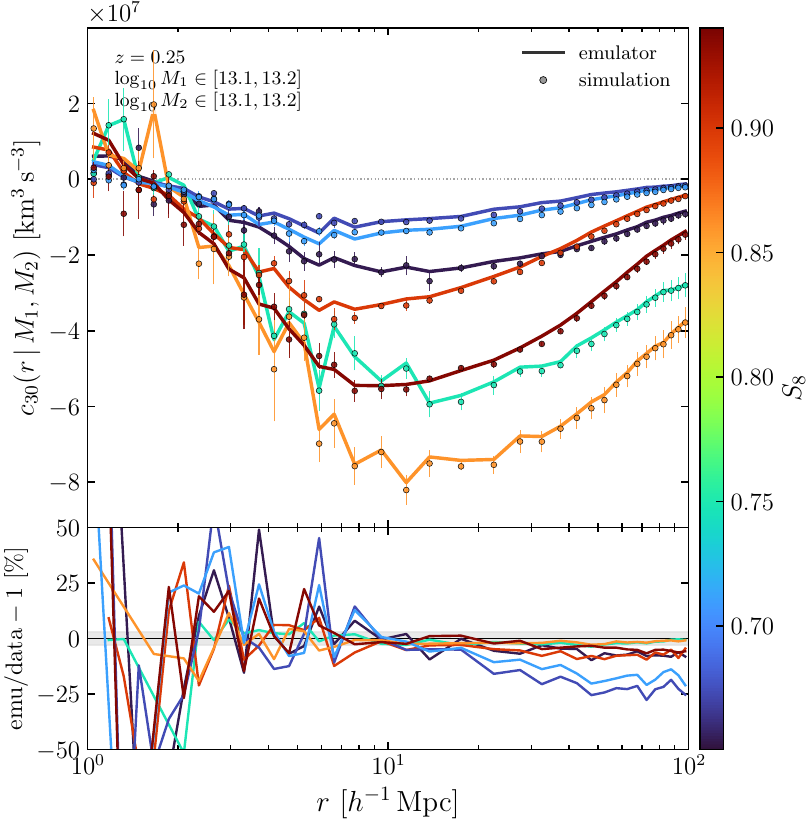}%
    \includegraphics[width=0.33\textwidth]{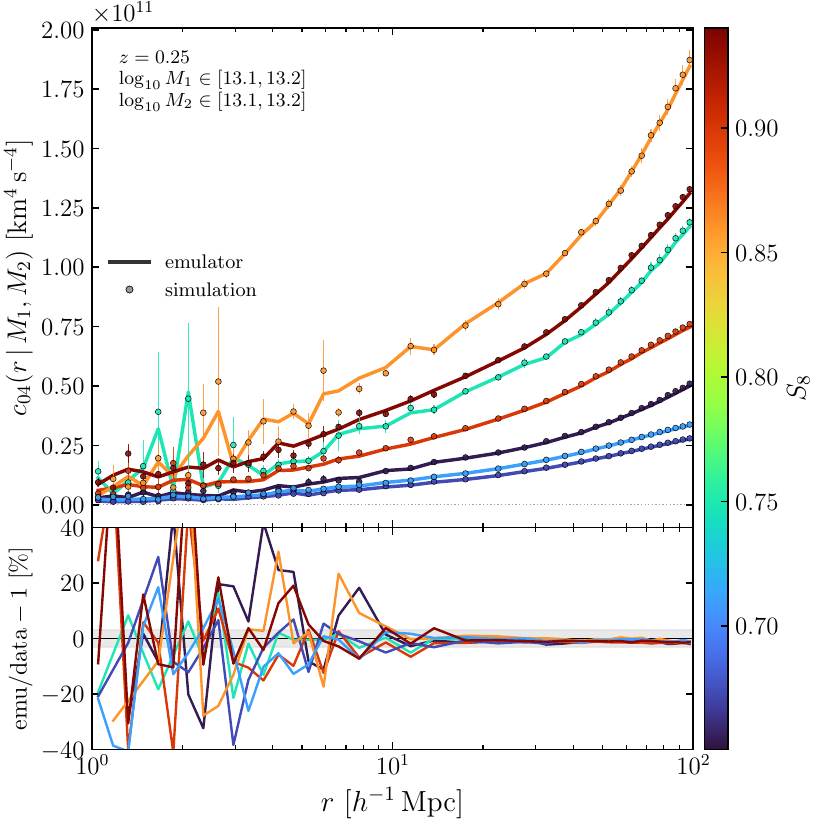}
    \caption{
        Emulator predictions for representative higher-order halo pairwise velocity moments:
        dispersion ($c_{02}$, left), skewness ($c_{30}$, centre), and kurtosis ($c_{04}$, right).
        Halo masses are fixed at $M_1=M_2=10^{13.1}\,h^{-1}M_{\odot}$ while colours denote \revision{the seven held-out test models~57--63, with the line colour keyed to $S_8$ via the colour bar}.
        Despite the increased noise present in differential mass-bin measurements, the emulator captures smooth scale and cosmology dependence across all moments.
        ~\href{https://github.com/chzruan/freyja/blob/main/paper_figs/halo_vm_plots.py}{\includegraphics[scale=0.019]{figs/freyja.jpg}}
    }
    \label{fig:halo_vm_c234}
\end{figure*}

Fig.~\ref{fig:halo_vm_c234} presents predictions for higher-order halo pairwise velocity moments, including the dispersion ($c_{02}$), skewness ($c_{30}$), and kurtosis ($c_{04}$). 
These quantities probe progressively non-Gaussian aspects of the pairwise velocity distribution and are therefore intrinsically noisier than the mean velocity. 
Nevertheless, after applying the physics-informed rescaling described above, the emulator recovers smooth and physically consistent trends with separation and cosmology. 
The residual fluctuations primarily reflect measurement noise in  differential mass bins and do not propagate significantly into halo-model predictions, where the moments enter through integrals over halo mass. 
This demonstrates that the adopted reduced parametrisation provides a stable basis for emulating the full hierarchy of velocity moments required by the streaming model.

\revision{Finally, we comment on the scalability of the Gaussian-process emulators. Each GP is trained on the $N_{\text{train}} = 54$ design cosmologies, and its dominant cost is the $\mathcal{O}(N_{\text{train}}^3)$ Cholesky factorisation of the training covariance matrix. For the present suite this cost is negligible compared with the neural-network training and the simulation cost. Were the training set to be enlarged to $\sim 500$--$1000$ cosmologies in pursuit of sub-percent precision, this $\mathcal{O}(N^3)$ scaling would eventually become a bottleneck; standard mitigations include sparse and inducing-point GPs \citep[e.g.][]{NIPS2005_4491777b,pmlr-v5-titsias09a}, product-of-experts ensembles, and deep-kernel methods. The modular design of the pipeline allows the back-end of any individual emulator to be replaced by one of these schemes, or by a neural network, without affecting the remainder of the pipeline.}


\section{Model Validation}
\label{sec:mcmc}

\begin{figure*}
    \centering
    \includegraphics[width=0.95\textwidth]{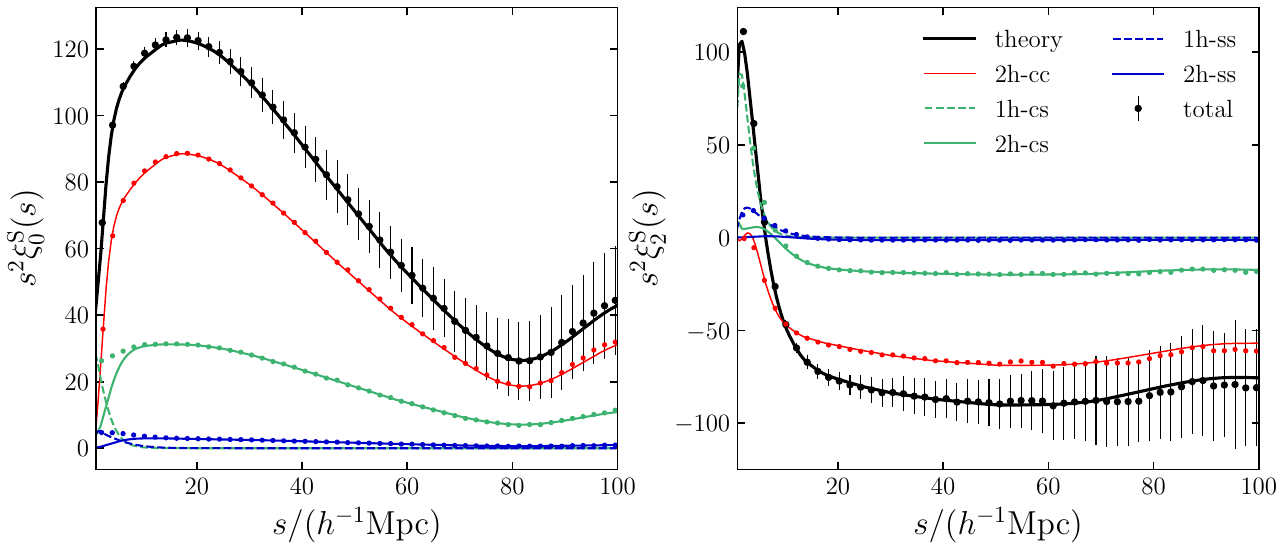}
    \caption{
        Predictions of the emulator-based halo streaming model for the galaxy redshift-space correlation function multipoles at $z=0.25$. \revision{The cosmology shown is held-out test model~$60$ of the suite, $(\Omega_{\text{m0}}, h, n_{\text{s}}, S_8) = (0.267, 0.733, 0.989, 0.823)$.}
        The left panel shows the monopole $\xi^{\text{S}}_0(s)$, while the right panel shows the quadrupole $\xi^{\text{S}}_2(s)$.
        Points denote measurements from simulations and curves show model predictions.
        Black lines correspond to the total galaxy-galaxy clustering signal.
        Contributions from different galaxy pair types are shown separately:
        central-central (red), central-satellite (green), and satellite-satellite (blue).
        Solid lines indicate two-halo terms and dashed lines indicate one-halo terms.        ~\href{https://github.com/chzruan/freyja/blob/main/paper_figs/plot_xiS024.py}{\includegraphics[scale=0.019]{figs/freyja.jpg}}
    }
    \label{fig:xiS}
\end{figure*}

Fig.~\ref{fig:xiS} shows the halo streaming predictions for the galaxy redshift-space distortion monopole and quadrupole, together with measurements from the simulations. \revision{These predictions correspond to held-out test model~$60$ of the suite, with $(\Omega_{\text{m0}}, h, n_{\text{s}}, S_8) = (0.267, 0.733, 0.989, 0.823)$, i.e.\ an $S_8$ near the centre of the training range.} 
The model accurately reproduces both multipoles across a broad range of scales, demonstrating that the emulator-based description of halo clustering and velocity statistics provides a consistent mapping from real to redshift space. 
The decomposition into galaxy pair types clarifies the physical origin of the signal: on large scales the monopole and quadrupole are dominated by the two-halo central-central contribution, reflecting coherent large-scale flows, while on  smaller separations the one-halo central-satellite and satellite-satellite terms become more important and drive the suppression of the quadrupole through virial motions. 
The agreement between data and model predictions across the transition between one- and two-halo regimes indicates that the halo streaming framework captures both linear Kaiser effects and nonlinear Fingers-of-God behaviour within a unified model.

To further validate the accuracy and internal consistency of this emulator-based halo streaming model, we perform an end-to-end parameter recovery test using Markov chain Monte Carlo  inference. The goal of this test is to assess whether the full modelling pipeline, including the emulators for halo properties, can recover the true cosmological and halo occupation distribution parameters.

\revision{The cosmology fitted in this recovery test is the fiducial \textit{Planck}-2015 $\Lambda$CDM cosmology of the simulation suite \citep{Planck15Parameters:2016A&A...594A..13P}, with $\Omega_{\text{m0}} = 0.3089$, $h = 0.6774$, $\sigma_8 = 0.8159$ (i.e.\ $S_8 = 0.828$) and $n_{\text{s}} = 0.9667$; it is marked with a star in Fig.~\ref{fig:cosmoparams}, where it lies close to the centre of the design. The fit uses the monopole and quadrupole in $32$ logarithmically spaced separation bins covering $1 < s \le 117\,h^{-1}\,\mathrm{Mpc}$ (i.e.\ $s_{\min} = 1\,h^{-1}\,\mathrm{Mpc}$, with the largest bin centred at $s \simeq 117\,h^{-1}\,\mathrm{Mpc}$), together with the galaxy number density assigned a $5$ per cent Gaussian relative uncertainty.}

We define the Gaussian likelihood as
\begin{align}
    \ln \mathcal{L}\qty(\boldsymbol{\theta}_{\text{cos}}, \boldsymbol{\theta}_{\text{HOD}} \,|\, \boldsymbol{d}_{\text{obs}}) 
    \propto 
    &\Big[ \boldsymbol{d}_{\text{obs}} - \boldsymbol{d}_{\text{model}}(\boldsymbol{\theta}_{\text{cos}}, \boldsymbol{\theta}_{\text{HOD}}) \Big]^{\intercal}
    \boldsymbol{C}^{-1} \notag \\
    \phantom{\propto} &\Big[ \boldsymbol{d}_{\text{obs}} - \boldsymbol{d}_{\text{model}}(\boldsymbol{\theta}_{\text{cos}}, \boldsymbol{\theta}_{\text{HOD}}) \Big],
\end{align}
where $\boldsymbol{d}_{\text{obs}}$ is the ``observed'' data vector and $\boldsymbol{d}_{\text{model}}(\boldsymbol{\theta}_{\text{cos}}, \boldsymbol{\theta}_{\text{HOD}})$ denotes the prediction of the halo streaming model for a given set of cosmological parameters $\boldsymbol{\theta}_{\text{cos}}$ and HOD parameters $\boldsymbol{\theta}_{\text{HOD}}$. 
\revision{The data vector is the \textit{mean} of the $100$ fiducial \textit{Planck}-cosmology realisations; averaging suppresses the measurement noise of the target itself (by a factor of $10$), so that $\boldsymbol{d}_{\text{obs}}$ provides an effectively noise-free benchmark against which residual model systematics can be assessed.}
The covariance matrix $\boldsymbol{C}$ is estimated from the AbacusSummit \citep{Maksimova:2021MNRAS.508.4017M.ABACUSSUMMITOverview} small ($500\,h^{-1}\,\mathrm{Mpc}$) box simulations and is kept fixed. 
\revision{We rescale the covariance matrix correspondingly to match the effective volume of the averaged data vector. The resulting covariance therefore represents the statistical uncertainty of the $100$-realisation mean and is used consistently in the likelihood when assessing parameter shifts. Establishing the systematic error floor at the full effective volumes of forthcoming surveys will require the corresponding survey-volume covariance together with tighter convergence of the emulators, and will be presented alongside the survey analyses in forthcoming data papers.}

\begin{figure*}
    \centering
    \includegraphics[width=0.95\textwidth]{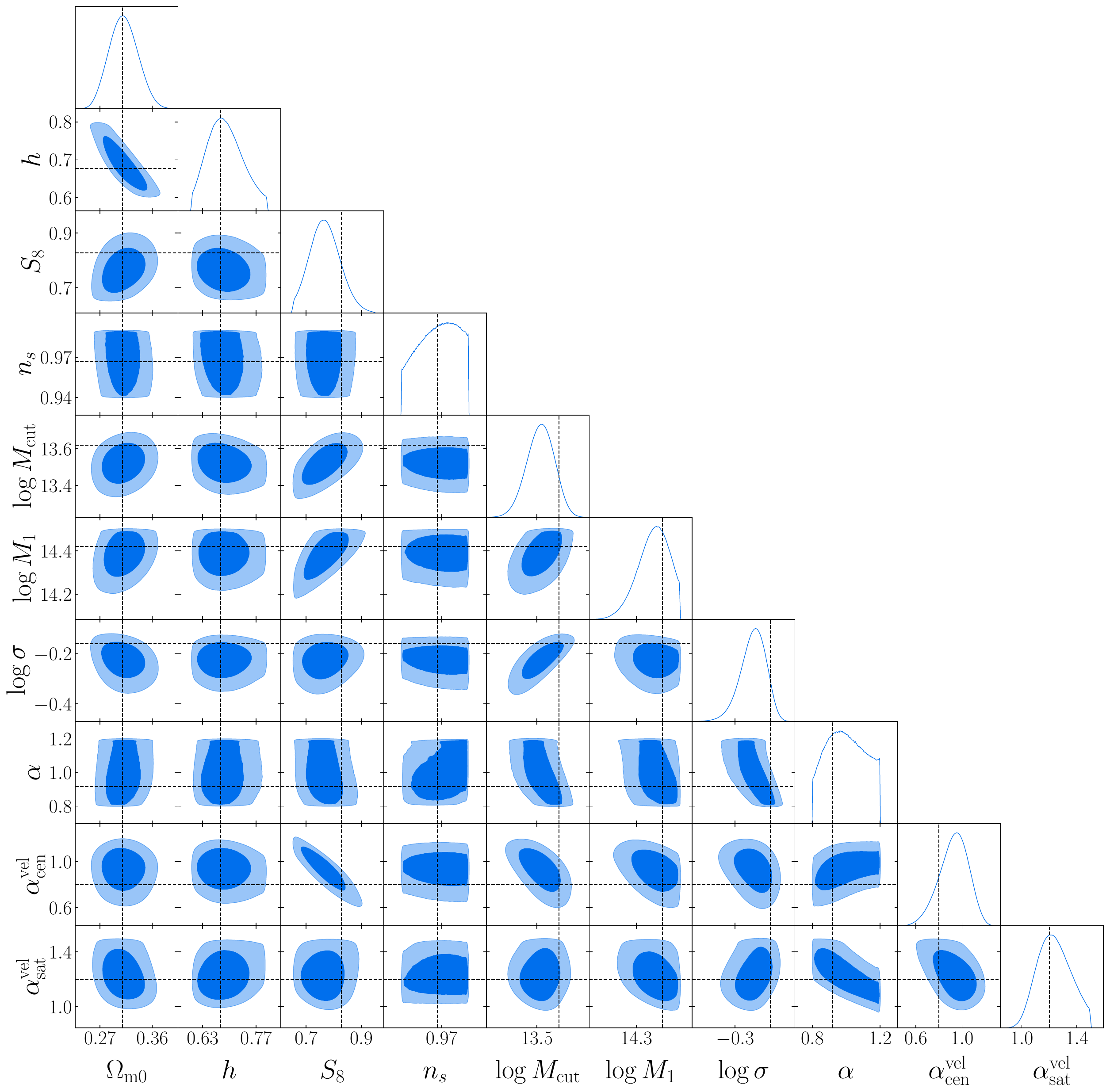}
    \caption{Corner plot from the MCMC recovery test using the halo streaming model.
    The contours show the joint and marginal posterior distributions of all cosmological and halo occupation distribution parameters, with the input values indicated by vertical and horizontal reference lines.
    All parameters are successfully recovered within the $1\sigma$ credible intervals, demonstrating that the emulator based model is unbiased and retains the cosmological information required for joint inference of cosmology and galaxy bias.
    \revision{The data vector is the mean of the $100$ fiducial \textit{Planck}-cosmology realisations. The fit combines the monopole and quadrupole over $1 < s \le 117\,h^{-1}\,\mathrm{Mpc}$ ($32$ bins) with the galaxy number density ($5$ per cent Gaussian uncertainty). The satellite and central fractions $f_{\text{sat}}$ and $f_{\text{cen}}$ are not free parameters and therefore do not appear here: they are derived from the HOD and cosmological parameters via the integral of the halo mass function against the occupation functions (see Section~\ref{sec:mcmc}). The marginal $\log M_1$ posterior leans against its prior edge because the satellite fraction is small and $\log M_1$ is weakly constrained and degenerate with the other occupation and velocity-bias parameters; this does not bias the recovered cosmology.}
    }
    \label{fig:mcmc_corner}
\end{figure*}

Fig.~\ref{fig:mcmc_corner} shows the posterior distributions from this likelihood. The contours indicate the joint and marginal constraints on all cosmological and HOD parameters, while the reference lines mark the true input values used to generate the mock data. All parameters are successfully recovered within the one sigma credible intervals, with no evidence for systematic shifts or biased degeneracy directions. \revision{For the cosmological parameters we recover $\Omega_{\text{m0}} = 0.31 \pm 0.02$ and $S_8 = 0.79 \pm 0.07$, to be compared with the input values $0.3089$ and $0.828$, i.e.\ shifts of $\lesssim 0.5\sigma$ relative to the adopted covariance.} This shows that the emulated halo streaming model provides an unbiased mapping between parameter space and observable space. \revision{The uniform priors adopted for all parameters are summarised in Table~\ref{tab:priors}.}

Overall, this recovery test confirms that the modular emulation framework preserves the cosmological and astrophysical information content of the clustering signal, while maintaining sufficient accuracy for joint inference. The successful parameter recovery provides a critical validation of the model and supports its application to precision analyses of spectroscopic galaxy surveys.

\revision{Throughout, the data points in the emulator-validation figures (Figs.~\ref{fig:hmf_emulation_wrfsdv}, \ref{fig:matter_pk_emulator_test}, \ref{fig:xi_hh_emu}, \ref{fig:bias_tinker_ext}, \ref{fig:xi_hh_emu} and \ref{fig:halo_vm_m10}) carry across-realisation standard-error-of-the-mean error bars, $\sigma_{\text{box}} / \sqrt{N_{\text{box}}}$, estimated from the five realisations per cosmology. This is exactly the per-point noise model used to weight the Gaussian-process training, and it is the reference against which the emulator residuals are gauged.}

\revision{We note that the central and satellite fractions $f_{\text{cen}}$ and $f_{\text{sat}}$ do not appear as sampled parameters in Fig.~\ref{fig:mcmc_corner}. They are not free parameters of the model but derived quantities, e.g.\ $f_{\text{cen}} = \bar n_g^{-1} \int \dd M\, (\dd n / \dd M)\, \langle N_{\text{c}}\rangle(M)$ and analogously for $f_{\text{sat}}$, with $\bar n_g$ the integrated mean galaxy number density. They are therefore fully determined by the HOD and cosmological parameters already shown and carry no additional freedom.}

\begin{figure}
    \centering
    \includegraphics[width=\columnwidth]{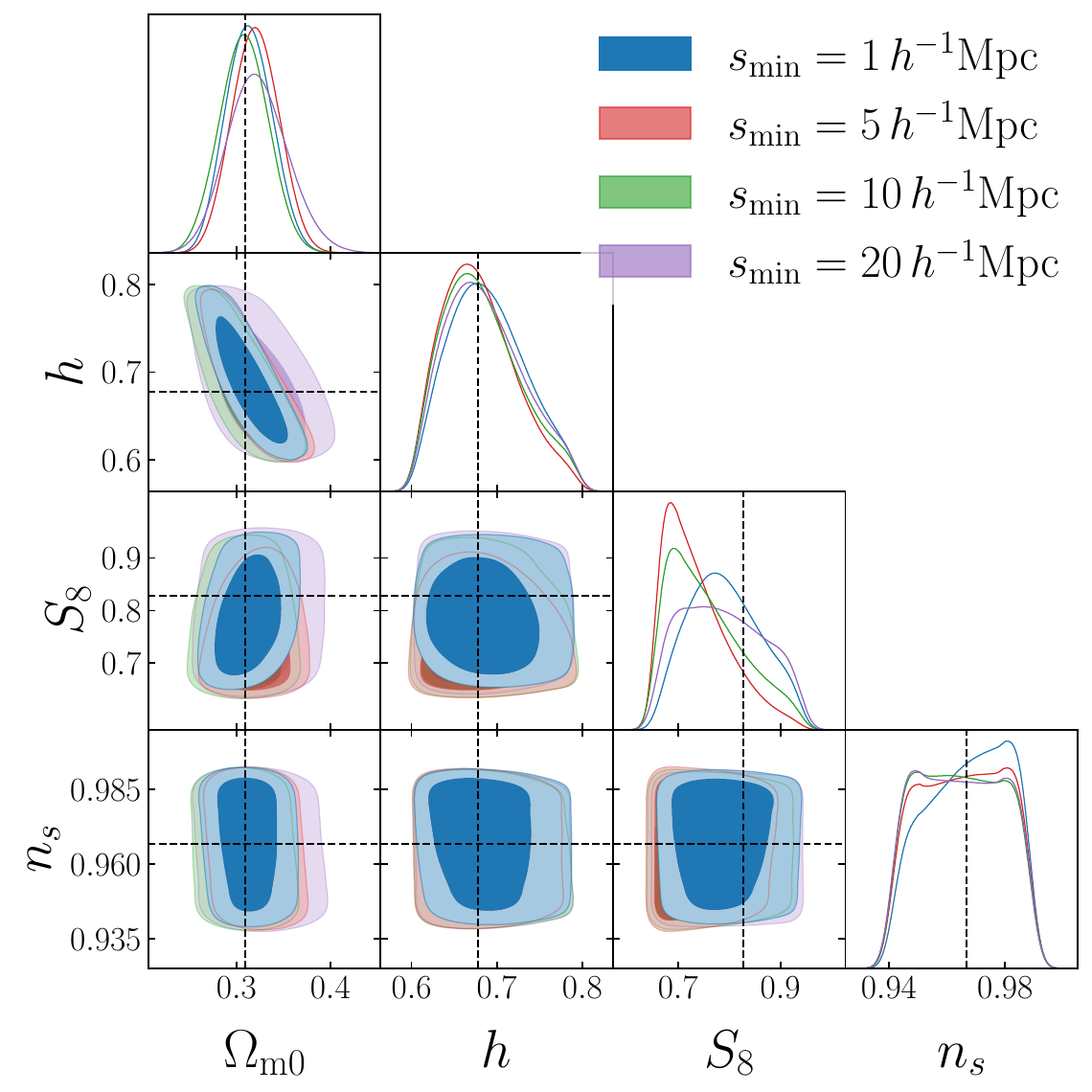}
    \caption{\revision{Robustness of the parameter recovery to the minimum fitted scale. The fit of Fig.~\ref{fig:mcmc_corner} is repeated for $s_{\min} \in {1, 5, 10, 20},h^{-1},\mathrm{Mpc}$ at fixed $s_{\max} \simeq 120,h^{-1},\mathrm{Mpc}$, fitting the monopole and quadrupole at $z=0.25$. Contours show the two-dimensional marginalised posterior distributions for the different scale cuts; reference lines mark the input parameter values. Most parameters remain stable across the tested scale cuts, although $S_8$ exhibits a residual dependence on $s_{\min}$. Including smaller scales generally tightens the statistical constraints.}}
    \label{fig:smin_evolution}
\end{figure}

\revision{To verify that the recovery is robust to the choice of minimum scale, we repeat the fit for several values of $s_{\min} \in \{1, 5, 10, 20\}\,h^{-1}\,\mathrm{Mpc}$ at fixed $s_{\max}$. Fig.~\ref{fig:smin_evolution} shows that the recovered parameters remain mutually consistent and unbiased with respect to the input values across all scale cuts, while the marginal constraints tighten monotonically as smaller scales are included. This demonstrates that the additional small-scale information captured by the halo streaming model is exploited without introducing detectable systematic shifts.}
\revisiontwo{Although most recovered parameters show little dependence on the minimum fitted scale, a residual dependence remains for $S_8$. Given the sensitivity of small-scale clustering to the fluctuation amplitude, this behaviour warrants particular attention when including these scales. Applications to observational data should therefore adopt a conservative minimum fitting scale, established through mock-based recovery tests that verify that any systematic shifts, especially in $S_8$, are sufficiently small relative to the statistical uncertainties.}

\revision{Some marginal posteriors extend to the prior boundaries. For the cosmological parameters and the parameters that enter the emulators this simply reflects the limited constraining power of the single mock data vector. The HOD parameter $\log M_1$, which does not enter any emulator and acts only through the analytic occupation functions, also leans against its upper prior edge. This is a genuine lack of constraining power rather than a model deficiency: for the LOWZ-like sample used here the satellite fraction is small, so the redshift-space multipoles depend on $\log M_1$ mainly through the satellite fraction and its effect is largely degenerate with the other satellite-occupation parameters ($\sigma$, $\alpha$) and with the velocity-bias parameters ($\alpha_{\text{cen}}^{\text{vel}}$, $\alpha_{\text{sat}}^{\text{vel}}$) that control the Fingers-of-God damping. Crucially, this truncation does not propagate into the recovered cosmological parameters, which are unbiased as reported above.}

\begin{table}
    \centering
    \revision{
    \caption{\revision{Uniform priors adopted in the parameter-recovery analysis of Section~\ref{sec:mcmc}. Cosmological parameters are listed in the upper block and HOD parameters in the lower block. The HOD parameter $\sigma$ is sampled in $\log_{10}$, so $\log\sigma \in [-0.699, -0.046]$ corresponds to $\sigma \in [0.20, 0.90]$.}}
    \label{tab:priors}
    \begin{tabular}{lc}
        \hline
        Parameter & Prior range \\
        \hline
        $\Omega_{\text{m0}}$ & $[0.15,\,0.45]$ \\
        $h$ & $[0.60,\,0.80]$ \\
        $S_8$ & $[0.65,\,0.95]$ \\
        $n_{\text{s}}$ & $[0.94,\,0.99]$ \\
        \hline
        $\log M_{\text{cut}}$ & $[12.8,\,13.8]$ \\
        $\log M_1$ & $[13.2,\,14.5]$ \\
        $\log\sigma$ & $[-0.699,\,-0.046]$ \\
        $\alpha$ & $[0.8,\,1.2]$ \\
        $\alpha_{\text{c}}^{\text{vel}}$ & $[0.5,\,1.5]$ \\
        $\alpha_{\text{s}}^{\text{vel}}$ & $[0.5,\,1.5]$ \\
        \hline
    \end{tabular}
    }
\end{table}

\section{Discussion and Conclusions}
\label{sec:conclusion}

We have developed an emulator-based halo streaming model that provides an accurate, interpretable, and computationally efficient description of nonlinear redshift-space distortions in galaxy clustering.
By combining the halo model with the streaming-model formalism and predicting each building block ingredient using emulators trained on $N$-body simulations, we construct a unified framework capable of predicting the full-shape anisotropic clustering signal across a wide range of scales and tracer populations.

A key aspect of our work is the principle of \textit{modular emulation}. 
Rather than directly emulating the       redshift space observables, which would require generating a large number of mock galaxy catalogues, $N_{\text{cosmology}} \times N_{\text{HOD}}$, we focus instead on the physical building blocks of the model, namely the halo mass function, real space halo correlation functions, and pairwise velocity moments. 
This strategy preserves physical transparency and interpretability while allowing each component to be optimized, validated, and extended independently. 
In practice, such modularity yields greater robustness and flexibility than end to end emulation approaches, particularly when varying tracer properties or incorporating additional physical effects.
Nevertheless, \revision{direct emulation approach \citep[e.g.][]{Zhai:2017ApJ...848...76Z,Kwan:2023ApJ...952...80K,Wang_2024,Wang:2025ApJ...994...51W,Zhai:2025MNRAS.544..932Z} remains valuable, and must be adopted} in cases where analytical decompositions are impractical, such as for void \citep[e.g.][]{Euclid:2023eom,Fraser:2025JCAP...06..001F} and density split statistics \citep[e.g.][]{Cuesta-Lazaro:10.1093/mnras/stae1234,Paillas:2023cpk}.

The halo streaming model systematically decomposes the RSD signal into contributions from different galaxy types and halo environments.
This provides clear physical insight into the origin of nonlinear RSD effects.
Small-scale Fingers-of-God features are primarily sourced by satellite motions within haloes, while large-scale anisotropies are governed by coherent halo bulk flows induced by gravity.
By explicitly modelling these components and their associated velocity moments, the framework captures the transition from linear to deeply nonlinear scales in a controlled and physically motivated manner.
The agreement with simulation-based mock catalogues demonstrates that this approach can recover the full-shape RSD signal without discarding valuable small-scale information.

An important advantage of the present framework is its computational efficiency.
Once trained, the emulators enable rapid predictions of all halo-model ingredients, allowing the full redshift-space correlation function to be evaluated without generating full HOD catalogues.
This efficiency makes the model well suited for exploring baseline extensions and performing cosmological parameter inference.
At the same time, the separation between cosmological dependence and tracer modelling also facilitates joint analyses of cosmology and galaxy-halo connection parameters.

Although the model achieves high accuracy within the regime explored in this work, several limitations and avenues for improvement remain.
First, the accuracy of the predictions is ultimately bounded by the fidelity and coverage of the training simulations.
Extensions to larger cosmological parameter spaces, different redshifts, or alternative gravity models will require additional simulation suites. \revision{We are in particular constructing a higher-resolution suite that will extend the reliably resolved halo-mass range to lower masses, enabling ELG- and BGS-targeted analyses in follow-up work.}
We plan to run suites of simulations for alternative gravity theories such as $f(R)$ gravity and Dvali-Gabadadze-Porrati (DGP) model.
Second, the current implementation assumes standard halo definitions and parametric HOD prescriptions; more flexible or assembly-biased galaxy-halo connections could be incorporated in future work. \revision{In particular, the one-halo pairwise velocity probability distributions (Section~\ref{subsec:pdf_vlos_model_skewt}) presently parameterise only the random component of the intra-halo satellite motions; a coherent satellite velocity bias --- a deterministic radial shift of the satellite velocity distribution that alters the first moment of the line-of-sight pairwise velocity PDF \citep[e.g.][]{Yuan:2024MNRAS.530..947Y} --- is a natural extension that can be accommodated within the modular framework at the cost of one additional emulator input.}
Third, baryonic effects are not explicitly incorporated and may impact clustering and velocity statistics on scales $r \lesssim 1$--$2\,h^{-1}\mathrm{Mpc}$, where processes such as gas cooling, star formation, and feedback from supernovae and active galactic nuclei can modify the internal structure of haloes and the distribution and velocities of galaxies relative to the dark matter \citep[e.g.][]{Mead:2015MNRAS.454.1958M,Chisari:2019OJAp....2E...4C}.

The halo streaming framework presented here is readily extensible.
The formalism can be applied to any galaxy tracer, three-point correlation function, or alternative tracers such as neutral hydrogen and quasars.
Incorporating modified gravity models and environmental dependencies of galaxy occupation are natural next steps.
Moreover, the modular structure makes it straightforward to replace or augment individual ingredients as improved simulations or theoretical models become available.


\revision{It is worth framing the current accuracy against the requirements of upcoming analyses. 
Every ingredient entering the halo streaming model is emulated as a physically normalised residual: for example, the matter sector is described as the departure from analytic linear theory, while the halo sector is represented through the dimensionless departure
$D = \xi_{\text{hh}} / [b(M_1) b(M_2) \xi_{\text{mm}}] \to 1$
from the linear-bias factorisation.
These reduced quantities vary smoothly across the sampled cosmologies, and the leave-one-out errors of the emulators are comparable to the across-realisation noise floor of the training simulations.
At the present simulation volume and for the scale range targeted here, the validation accuracy is therefore limited primarily by finite-volume and realisation noise rather than by cosmological interpolation within the training set.
On held-out cosmologies, the cumulative halo mass function, $P_{\text{mm}}(k)$, and $\xi_{\text{mm}}(r)$ are emulated with $0.3$--$0.5$ per cent root-mean-square accuracy, while the binned $\xi_{\text{hh}}$ is reproduced at the $1$--$2$ per cent level below $30\,h^{-1}\,\mathrm{Mpc}$.
This level of performance is adequate for the intermediate-scale measurements targeted in this work.
Reaching the sub-percent matter-sector precision required on the smallest scales by forthcoming DESI and \textit{Euclid} analyses will nevertheless require both denser sampling of cosmological parameter space and lower-noise simulations, including additional training nodes and larger simulation volume.
Both extensions are planned, and the modular design also allows external matter and halo emulators to be ingested directly.}

\revisiontwo{The upcoming work will extend the halo streaming model framework to incorporate environment-aware galaxy assembly bias~\citep[e.g.][]{Hadzhiyska:2020MNRAS.493.5506H,Yuan:2021MNRAS.502.3582Y,Collier:2026MNRAS.551g1559C}. We will model the galaxy-halo connection using a decorated HOD, allowing galaxy occupation to depend not only on halo mass but also on secondary halo properties and local environment density. In parallel, we are developing emulators for environment-dependent halo properties, including halo clustering and pairwise velocity statistics, so that these effects can be propagated consistently into the theory predictions. This extension will enable a more flexible and physically motivated treatment of assembly bias within the framework.}

In summary, we have demonstrated that combining simulation-based calibration with a physically motivated halo streaming model provides a powerful route toward precision modelling of nonlinear redshift-space distortions.
This approach bridges the gap between analytic theory and numerical simulations, retaining interpretability while achieving the accuracy required by forthcoming surveys.
With its flexibility, efficiency, and transparent physical foundation, the framework developed in this work offers a robust tool for extracting cosmological information from the full statistical power of next-generation large-scale structure data sets such as DESI and \textit{Euclid}.




\section*{Acknowledgements}
We thank the referee for their careful reading of the manuscript and for constructive comments that helped improve its clarity and presentation.
C-ZR and BL are funded by the ERC Advanced Grant UNCA (UKRI Frontiers Research Guarantee No.~EP/Z533877/1). BL, CMB and SB acknowledge support from the UK STFC Consolidated Grant ST/X001075/1. C-ZR and DFM thank the Research Council of Norway for their support and the resources provided by UNINETT Sigma2 -- the National Infrastructure for High-Performance Computing and Data Storage in Norway. S.B. is supported by the UKRI Future Leaders Fellowship [grant numbers MR/V023381/1 and UKRI2044].
This work used the DiRAC@Durham facility managed by the Institute for Computational Cosmology on behalf of the STFC DiRAC HPC Facility (www.dirac.ac.uk). The equipment was funded by BEIS capital funding via STFC capital grants ST/K00042X/1, ST/P002293/1, ST/R002371/1 and ST/S002502/1, Durham University and STFC operations grant ST/R000832/1. DiRAC is part of the National e-Infrastructure.

\section*{Data Availability}

The data underlying this article will be shared on reasonable request to the corresponding author.
The processed data and plotting scripts are available in the \href{https://github.com/chzruan/freyja/tree/main/paper_figs}{\texttt{freyja}} package. 
We also provide direct access to the source code for each figure via the clickable icon appended to the figure captions.





\bibliographystyle{mnras}
\bibliography{halostreaming}



\appendix


\bsp	
\label{lastpage}
\end{document}